\documentclass[fleqn,usenatbib]{mnras}

\usepackage{newtxtext,newtxmath}
\usepackage{multirow}

\newcommand{\herschel}{\textit{Herschel}}

\newcommand{\magphys}{\textsc{magphys}}

{}

{}

{}

{}

\usepackage[T1]{fontenc}
\usepackage{comment}

\defcitealias{Dunne2011}{D11}
\defcitealias{Beeston2018}{B18}
\defcitealias{Page2000}{PC00}
\DeclareRobustCommand{\VAN}[3]{#2}
\let\VANthebibliography\thebibliography
\def\thebibliography{\DeclareRobustCommand{\VAN}[3]{##3}\VANthebibliography}

\usepackage{graphicx}	
\usepackage{amsmath}	

\title[Evolution of the Dust Mass Function]{Confirming the Evolution of the Dust Mass Function in Galaxies over the past 5 Billion Years }

\author[R. A. Beeston, H.L. Gomez et al.]{
R. A Beeston,$^{1}$
H.L. Gomez,$^{1}$\thanks{E-mail: gomezh@astro.cf.ac.uk}
L Dunne,$^{1}$
S Maddox,$^{1}$
S. A. Eales$^{1}$
and M. W. L. Smith,$^{1}$
\\
$^{1}$Cardiff Hub for Astrophysical Research and Technology (CHART), School of Physics \& Astronomy, Cardiff University, The Parade, Cardiff, CF24 3AA, UK
}

\date{Accepted XXX. Received YYY; in original form ZZZ}

\pubyear{2015}

\begin{document}
\label{firstpage}
\pagerange{\pageref{firstpage}--\pageref{lastpage}}
\maketitle

\begin{abstract}
The amount of evolution in the dust content of galaxies over the past five billion years of cosmic history is contested in the literature. Here we present a far-infrared census of dust based on a sample of 29,241 galaxies with redshifts ranging from $0 < z < 0.5$ using data from the {\it Herschel Astrophysical Terahertz Survey} ($H$-ATLAS). We use the spectral energy distribution fitting tool \magphys~and a stacking analysis to investigate the evolution of dust mass and temperature of far-infrared-selected galaxies as a function of both luminosity and redshift.  At low redshifts, we find that the mass-weighted and luminosity-weighted dust temperatures from the stacking analysis both exhibit a trend for brighter galaxies to have warmer dust. In higher redshift bins, we see some evolution in both mass-weighted and luminosity-weighted dust temperatures with redshift, but the effect is strongest for luminosity-weighted temperature. 
The measure of dust content in galaxies at $z<0.1$ (the Dust Mass Function) has a different shape to that derived using optically-selected galaxies from the same region of sky.  We revise the local dust mass density ($z<0.1$) to $\rho_{\rm d} =(1.37\pm0.08)\times 10^5 {\rm\,M_{\odot}\,Mpc^{-3}}\,h_{70}^{-1}$; corresponding to an overall fraction of baryons
(by mass) stored in dust of $f_{\rm mb}
{(\rm dust)} = (2.22\pm 0.13) \times 10^{-5}$.
We confirm evolution in both the luminosity density and dust mass density over the past few billion years ($\rho_{\rm d} \propto (1+z)^{2.6 \pm 0.6}$), with a flatter evolution than observed in previous FIR-selected studies. We attribute the evolution in $\rho_L$ and $\rho_m$ to an evolution in the dust mass.
\end{abstract}

\begin{keywords}
galaxies: general — galaxies: ISM — dust, extinction — infrared: galaxies — submillimetre: galaxies
\end{keywords}



\section{Introduction}
The past 8 billion years is one of the most interesting periods of galaxy evolution as it encompasses the strongest decline in the amount of star formation in the universe, where the largest morphological transformation of galaxies takes place. As cold gas is the fuel for star formation, and thus the driver of this decline, a census of the cold gas in a large representative sample of galaxies over cosmic time is vital for understanding how the ‘lights turned out’. Simulations predict very little-to-no evolution of the cold interstellar medium in galaxies in this crucial epoch \citep{Schaye2015,Diemer2019,Dave2020}. Yet, we currently lack a robust observational measure of the interstellar content in galaxies over this redshift range in order to test these predictions. One tracer of the cold interstellar medium of galaxies is the dust emission \citep{Eales2012,Scoville2016}, hence the statistical FIR luminosity function (LF) - an important tool for probing the obscured star formation - and the related dust mass function (DMF) - a direct measurement of the space density of galaxies as a function of dust mass - are important to measure. 

Previous studies have shown that the dust luminosity of  galaxies  appears  to  evolve  rapidly  with  redshift (e.g. \citealt{Huynh2007,Dye2010,Hwang2010,Marchetti2012,Patel2013,Gruppioni2013}).  For example using the $H$-ATLAS Science Demonstration Phase (SDP), \cite{Dye2010} found that the evolution of the 250\,\micron~luminosity density is proportional to $(1+z)^{7.1}$ out to a redshift of $\sim 0.2$. The driving force behind dust luminosity evolution in this paradigm has been attributed to the increased heating of the dust due to the higher star formation rates in the past \citep{Magdis2012,Rowan-Robinson2012,Symeonidis2013,Berta2013,Casey2014}. Alternatively, a decrease in dust mass in galaxies with time could be responsible, we know that the gas content of the Universe has been decreasing with time, and  typically  a  galaxy's  gas  mass  is  strongly  linked  to  its dust content (e.g. \citealt{Eales2012,Genzel2015,Scoville2016,Scoville2017,Tacconi2018,Millard2021}).  

 \citet{Dunne2011} (hereafter \citetalias{Dunne2011}) found no evidence for the evolution of dust temperature with either redshift or luminosity, based on the $H$-ATLAS Science Demonstration Phase (SDP) sample of $\sim$1800 galaxies selected in the FIR. Instead, they found that the dust mass has decreased rapidly (by a factor of five) over the past five billion years of cosmic history. They found a redshift-dependent relationship for dust density $\rho_d$\footnote{The \citet{Dunne2011} dust densities were multiplied by a factor of 1.4 to account for the known under density of the SDP field. We also note that since the SDP analysis, \herschel\ calibration factors have changed by 10-20\,per\,cent.} where $\rho_{\rm d} (z) \propto (1+z)^{4.5}$ to $z=0.35$ and $\rho_{\rm d} (z=0) = (0.98 \pm 0.14) \times 10^{5}\,{\rm M_{\odot}\,Mpc}^{-3}$. 

The dust mass density is derived by integrating the dust mass function (DMF), a measure of the space density of dust in galaxies as a function of dust mass.  Constraining the DMF is becoming more relevant given the widespread use of dust emission as a tracer for the gas mass of galaxies in recent years (\citealt{Eales2010,Eales2012,Magdis2012,Scoville2014,Scoville2017,Millard2021}; see also the comprehensive review of \citealt{Casey2014}). This is of particular interest given difficulties in observing atomic and molecular line gas mass tracers out to higher redshifts \citep{Tacconi2013,Catinella2015,Genzel2015,Dunne2021}. Ground and Balloon-based studies led to the the DMF being measured locally \citep{Dunne2000,Vlahakis2005} and at redshifts 1 and 2.5  \citep{Dunne2003,Eales2009}. Unfortunately these studies were hampered by small number statistics and difficulties with observing from the ground.

The advent of \herschel\ \citep{Pilbratt2010} and the \textit{Planck} Satellite revolutionised studies of dust in galaxies, as they enabled greater statistics, better sensitivity, wider wavelength coverage and the ability to observe orders of magnitude larger areas of the sky than possible before. A 250-$\mu$m selected dust mass function was created from 1867 galaxies out to redshift 0.5 \citep{Dunne2011}. Subsequently, \citet{Negrello2013} and \citet{Clemens2013} published the DMF of 234 local star-forming galaxies from the all sky \textit{Planck} catalogue. \citet{Clark2015} then derived a local DMF from H-ATLAS (a 250-$\mu$m selected sample consisting of 42 sources). showing that FIR selected surveys pick up dust-rich galaxies with colder dust temperatures than those selected at optical or near IR wavelengths. In \cite{Beeston2018} (hereafter \citetalias{Beeston2018}), we presented the dust properties of the $H$-ATLAS GAMA equatorial fields using a sample comprised of $\sim$16,000 optically-selected galaxies within the redshift range $0.002\le z \le 0.1$. Our DMF showed fewer galaxies with high dust mass than predicted by semi-analytic models and more galaxies with high dust mass than predicted by hydrodynamical cosmological
simulations. Neither suite of simulations could reproduce the observed DMF at redshifts below 0.1. 

Later \citet{Driver2018} produced a dust mass function and measure of the dust mass density out to a redshift of $z<5$ based on an optically-selected sample of hundreds of thousands of galaxies. They found a peak in the dust mass density at $z\sim 1$ ($\sim$ 8 billion years ago) potentially coinciding with the so-called peak epoch of star formation (see also \citealt{Cucciati2012,Burgarella2013}). However, they found no evidence that the dust content of galaxies was evolving in recent cosmic history (ie $z<0.5$), finding a relatively flat dust mass density with redshift in contrast to \citet{Dunne2011}. The lack of evolution in dust density in \citeauthor{Driver2018} would require a strong dust temperature evolution with redshift in order to explain the evolution in the dust luminosity.  

More recently, \cite{Pozzi2020} used a 160$\mu$m-selected catalogue of $\sim$5300 galaxies in the COSMOS field to estimate the DMF for galaxies up to high redshifts ($z \sim 2.5$). Although they found a peak in the dust mass density at similar redshifts to \citet{Driver2018}, they disagreed at redshifts $<0.5$, finding a clear positive trend with redshift in agreement with \citeauthor{Dunne2011}. It is unclear whether the differences in the observed trends are due to sample selection i.e. FIR- versus optically-selected sources, survey area (with larger surveys having smaller cosmic variance errors) or something else. The cause of the strong evolution in dust luminosity and the nature of the evolution of the dust content in galaxies in recent cosmic history still remain controversial (see the review by \citealt{Peroux2020}). 

In this paper, we study the dust content of galaxies taken from the same region of sky as \citetalias{Beeston2018}, but through a catalogue formed of galaxies selected on their FIR emission rather than their stellar content. We derive the dust masses for these sources using two methods. Firstly, we use the method outlined in \cite{Dunne2011}, which relies upon using the spectral energy fitting code \magphys~\citep{daCunha2008}. Secondly, we perform a stacking analysis similar to \cite{Bourne2012}, where we stack the \herschel~luminosities of galaxies we assume to have similar FIR SEDs in order to search for trends in dust properties with redshift and luminosity. We use both methods to derive dust mass functions (DMFs) in five redshift slices out to $z=0.5$. We test whether selecting galaxies on their dust content (FIR emission, eg this work, \citetalias{Dunne2011}, \citealp{Millard2021}) rather than for their stellar population (eg \citetalias{Beeston2018}, \citealp{Driver2009}) introduces systematic differences in the DMF at different epochs. We also test whether we can reproduce the rapid change in the dust content of galaxies over the past 5 billion years, using a larger sample.
Throughout this work we assume a cosmology of $\Omega_m = 0.3$, $\Omega_{\Lambda} = 0.7$ and $H_0 = 70\,\rm km\,s^{-1}\,Mpc^{-1}$.

\section{The Data}

\subsection{The Sample}
The FIR and sub-mm imaging data used to derive dust masses in this work are provided via the $H$-ATLAS\footnote{http://www.h-atlas.org/} \citep{Eales2010} DR1 sample. $H$-ATLAS is the largest extragalactic Open Time survey using \herschel~spanning $\sim$ 660 square degrees of sky with 600 hours of observations in parallel mode across five bands (100 and 160\,$\mu$m with PACS - \citealt{Poglitsch2010}, and 250, 350, and 500\,$\mu$m with SPIRE - \citealt{Griffin2010}). $H$-ATLAS was specifically designed to overlap with other large area surveys such as SDSS and GAMA\footnote{http://www.gama-survey.org/}.  The GAMA survey is a panchromatic compilation of galaxies built upon a highly complete magnitude limited spectroscopic survey of around 286 square degrees of sky (with limiting magnitude $r_{\rm petro} \leq 19.8\thinspace$mag). As well as spectrographic observations, GAMA has collated broad-band photometric measurements in up to 21 filters for each source from ultraviolet (UV) to FIR/sub-mm \citep{Driver2016,Wright2017}. The imaging data required to derive photometric measurements come from the compilation of many other surveys: GALEX Medium Imaging Survey \citep{Bianchi1999}; the SDSS DR7 \citep{Abazajian2009}, the VST Kilo-degree Survey (VST KiDS, \citealp{deJong2013}); the VIsta Kilo- degree INfrared Galaxy survey (VIKING, \citealp{deJong2013}); the Wide-field Infrared Survey Explorer (WISE, \citealp{Wright2010}); and the $H$-ATLAS \citep{Eales2010}. Here we use galaxies in the three equatorial fields of the GAMA survey (G09, G12 and G15), covering $\sim 180$ square degrees of sky between them. The GAMA/$H$-ATLAS overlap spans approximately 145\,sq.\,degrees.

 Photometry in the five \herschel~bands for the $H$-ATLAS DR1 is provided in \cite{Valiante2016} based on final \herschel\ maps described in full in \cite{Smith2017}. Sources were selected initially at 250\,$\mu$m using \textsc{madx} with $S/N > 4$ in any of the three SPIRE bands. In theory the selection of galaxies from the FIR maps could depend on any of the three SPIRE bands, but in practice this selection is mostly determined by the 250$\,$\micron~flux. 
 \citet{Bourne2016} identified optical counterparts to the $H$-ATLAS sources from the GAMA catalogue using a likelihood ratio technique \citep{SmithD2011}, we set the likelihood ratio reliability to $R>0.8$ and we flag quasars.  The final sample in this work consists of 29,241 galaxies at $z \le 0.5$. A high portion of galaxies in this sample (75\,per\,cent) have only one \herschel~band with a $>4\sigma$ measurement.

\subsection{Redshifts}

\begin{figure}
 \includegraphics[trim=0mm 10mm 0mm 0mm clip=true,width=\columnwidth]{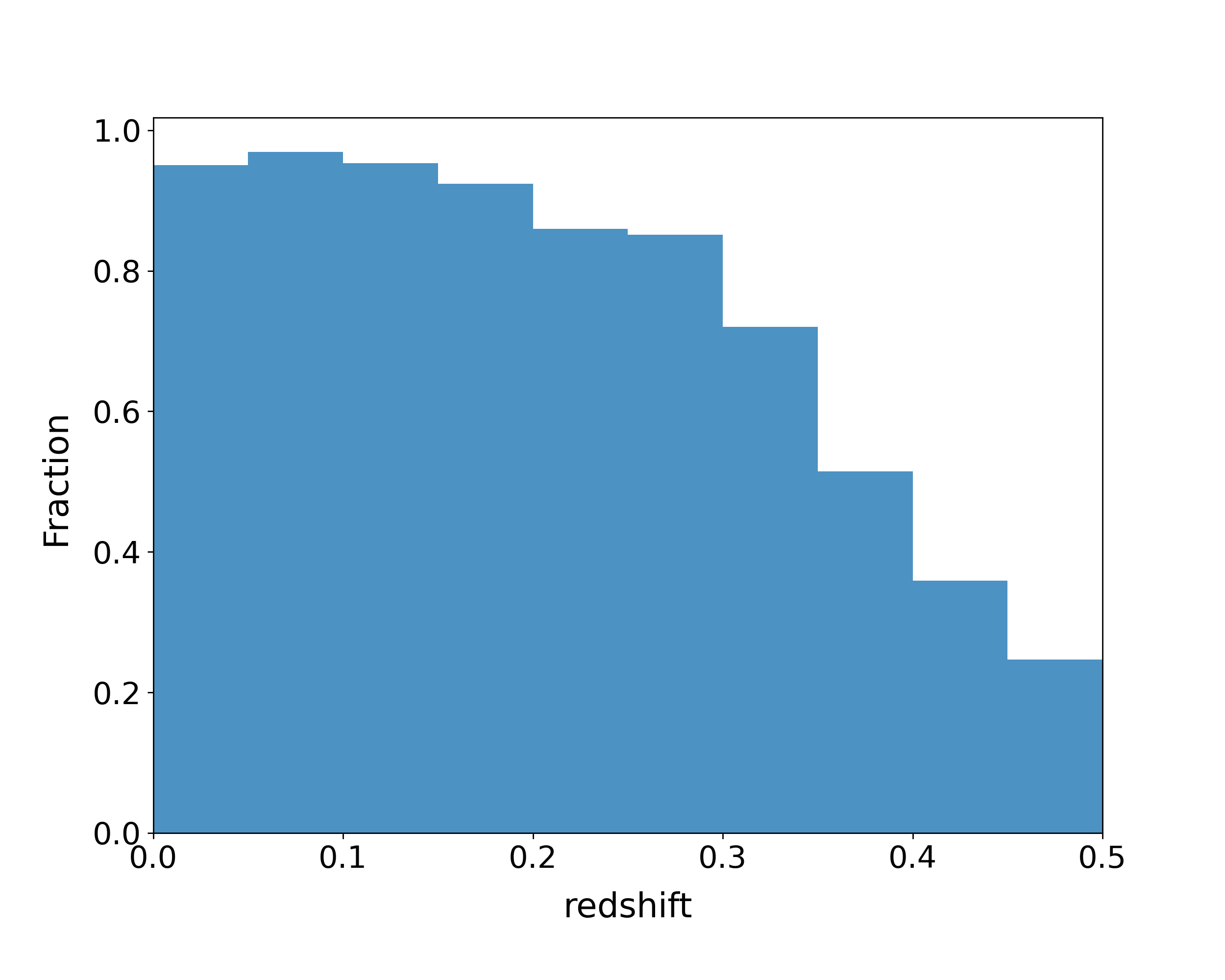}
 \caption{The fraction of galaxies in the sample with spectroscopic redshifts.}
 \label{fig:phot_redshifts}
\end{figure}

Where available, spectroscopic redshifts are used. GAMA compiled a catalogue of supplementary spectroscopic redshifts from SDSS DR7 and DR10 \citep{Ahn2013}, WiggleZ \citep{Drinkwater2010}, 2SLAQ LRG and QSO samples \citep{Cannon2006,Croom2009}, 6dF \citep{Jones2009}, MGC \citep{Driver2005}, 2QZ \citep{Croom2004}, 2dFGRS \citep{Colless2001}, and UZC \citep{Falco1999}. Further spectroscopic redshifts were provided for this work from the \herschel~Extragalactic Legacy Project (HELP, \citealp{Oliver2012,Shirley2021}). $H$-ATLAS did produce their own photometric redshifts using ANNz, an artificial neural network \citep{Collister2004}; however, these were shown to be biased beyond a redshift of $\sim 0.3$ by \citetalias{Dunne2011}. For those galaxies with no spectroscopic redshift, we use photometric redshifts from the Kilo Degree Survey (KiDS) catalogue \citep{deJong2017}. There are two different estimates for the redshift available from KiDS, firstly using an artificial neural network ANNz2 (the successor to ANNz, \citealp{Sadeh2016}), and secondly the Multi Layer Perceptron with Quasi Newton Algorithm (MLPQNA; \citealt{Bilicki2018}). A comparison of the redshift estimates from the available sources is provided in Appendix~\ref{app:redshift}.  Generally the photometric redshifts from ANNz calculated by $H$-ATLAS are in good agreement with both sets of redshifts from KiDS, albeit with a small tendency for $H$-ATLAS to under-estimate the redshift compared to KiDS at higher $z$. We choose to use the MLPQNA photometric redshifts for galaxies without a spectroscopic redshift in this work ($N \sim 7000$). The fraction of sources with spectometric redshifts is given in Figure~\ref{fig:phot_redshifts} and Table~\ref{tab:redshifts}.

\begin{table}
\centering
\begin{tabular}{ccccc}
 \hline
\multicolumn{1}{c}{Redshift} & \multicolumn{2}{c}{spec $z$} & \multicolumn{2}{c}{\magphys} \\
 \multicolumn{1}{c}{} & \multicolumn{1}{c}{$N$} & \multicolumn{1}{c}{$F$} & \multicolumn{1}{c}{$N$} & \multicolumn{1}{c}{$F$} \\ \hline
  0.0 - 0.1        &  3609    &   0.96 & 3454 & 0.93 \\
  0.1 - 0.2        &   7265    &   0.94 & 7095 & 0.92\\
  0.2 - 0.3        &   5443    &  0.86 & 5400 &0.85\\
  0.3 - 0.4      &   3609    &    0.62  & 3604 &0.62\\
  0.4 - 0.5      &  1684    &   0.30 & 1617 &0.29\\ \hline
\end{tabular}
\caption{The number $N$ and fraction $F$ of galaxies with spectroscopic redshifts in each redshift slice. The number of sources with \magphys~fits are shown in the last column.
}
\label{tab:redshifts}
\end{table}

\subsection{Calculating Luminosities}

Monochromatic rest-frame luminosities are calculated using the \herschel~fluxes derived by the $H$-ATLAS consortium \citep{Valiante2016}, and the equation:

\begin{equation}
L_{\nu} = \frac{4\pi D_{\rm L}^{2} S_{\nu} k}{(1+z)}.
\end{equation}

$L_{\nu}$ and $S_{\nu}$ are the luminosity and flux respectively at frequency $\nu$, $D_{\rm L}$ is the luminosity distance, $z$ is the redshift of the source, and $k$ is the $k$-correction. This is given by:

\begin{equation}
k = \left(\frac{\nu_{\rm rest}}{\nu_{\rm obs}}\right)^{3+\beta} \frac{e^{h\nu_{\rm obs}/k_{\rm B}T_d} - 1}{e^{h\nu_{\rm rest}/k_{\rm B}T_d} - 1},
\end{equation}

\noindent where $\nu_{\rm obs}$ and $\nu_{\rm rest}$ are the observed and rest-frame frequency respectively, $k_{\rm B}$ is the Boltzmann constant, $\beta$ is the dust emissivity index, and $T_d$ is the dust temperature.  In deriving the luminosity function here (and in order to compare with previous works), we follow \citetalias{Dunne2011} and \citetalias{Beeston2018} and correct for the field-by-field density variations in the GAMA fields at $z<0.1$. The fields are corrected by a factor of 1.36, 1.22, and 0.98, to account for the underdensity in regions G09 and G15, and the overdensity in G12 respectively, \citealt{Wright2017}).

\section{Estimating Dust Properties}
\label{sec:dustprops}

We perform two methods of estimating dust properties for the galaxies in the $H$-ATLAS DR1 fields. First we fit IR-submm SEDs of individual galaxies (Section~\ref{sec:HATLAS_magphys}), and second we put galaxies into redshift and luminosity bins and fit their stacked SEDs (Section~\ref{sec:stacking}). 

\subsection{\magphys~estimates of Dust Properties}
\label{sec:HATLAS_magphys}

The SED-fitting tool \magphys~\citep{daCunha2008} was applied to a subsection of the $H$-ATLAS DR 1 galaxies \citep{Eales2018} at redshifts $<0.5$ using the method outlined in \citet{Smith2012}. In the first instance we only fit galaxies with a \emph{spectroscopic redshift}.

\magphys~fits model SEDs to galaxy spectra using $\chi^2$ minimisation and vast libraries of optical and FIR models. \magphys~requires energy balancing, so that all the energy absorbed in the optical regime must be re-radiated in the FIR regime.  It returns various stellar and dust properties, returning both a `best fit' for each parameter as well as percentile values corresponding to the median, and the 1, 2, and 3$\,\sigma$ confidence intervals. 
The number of galaxies with a dust mass derived using \magphys~is 21,187 galaxies i.e. $\sim$72\,per\,cent of the sample.

\subsubsection{Deriving a Relationship Between 250$\, \mu m$ Luminosity and Dust Mass}

\label{sec:pin_magphys}
\begin{figure}
\includegraphics[trim=2mm 15mm 0mm 0mm clip=true,width=\columnwidth]{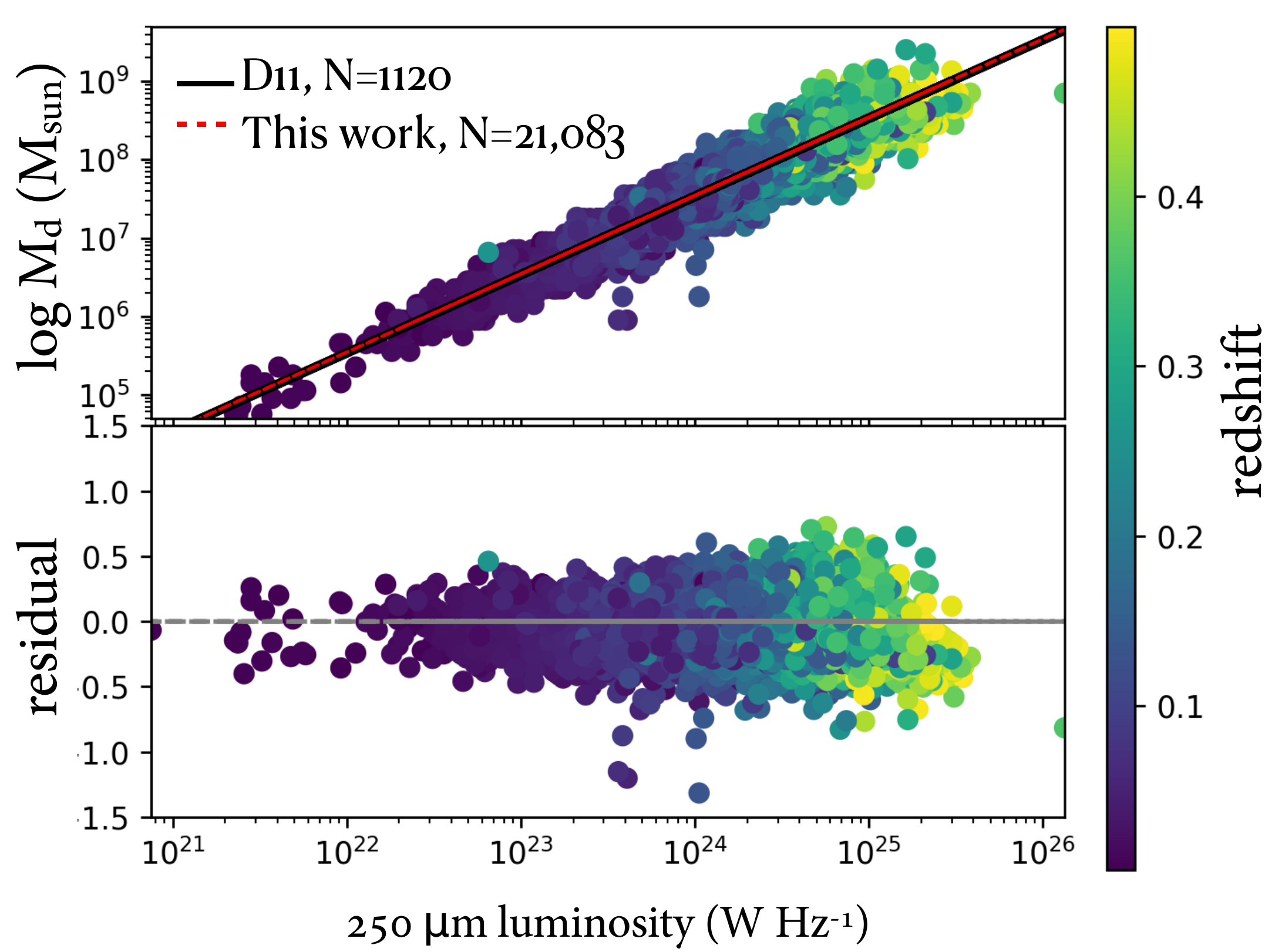}
 \caption{{\it Top:} The best-fit dust masses and luminosities of 21,187 galaxies from the $H$-ATLAS DR1 with a spectroscopic redshift $<0.5$ derived using \magphys~\citep{Eales2018}. A linear relationship is fit to the data (red dashed line). The black line shows the previous fit to the $\sim$1000 $\it H$-ATLAS SDP field  \citetalias{Dunne2011}. {\it Bottom:} A residual plot indicating the spread of the data which is large at redshifts greater than 0.35.}
 \label{fig:D11_mass_pin}
\end{figure}

To obtain a dust mass estimate for the remaining $\sim$ 30\,per\,cent of sources not fit by \magphys, we scale the 250\,\micron~luminosities ($L_{250}$) following the method described in \citetalias{Dunne2011}. They used a sample of 1120 galaxies from $H$-ATLAS in the SDP field to show that the observed $L_{250}$ and dust mass can be described by a simple linear relationship,

\begin{equation}
{\rm log}(M_d) = {\rm log}(L_{250}) - C,
\label{eq:linear_D11}
\end{equation}

\noindent with $C=16.47$. Here, we carry out the same analysis on a sample $\sim$ 20 times larger, an order of magnitude increase in the sky area and two orders of magnitude increase in the 250\,\micron~luminosity range. We derive the same value ($C=16.46$, Figure \ref{fig:D11_mass_pin}). We use this relationship to estimate the dust masses for the galaxies without spectroscopic redshifts. Although this may introduce bias into our results since the galaxies used to derive the relationship will be those that are optically bright enough to have a spectroscopic redshift. However using this relationship is, in effect, the same as applying a one component MBB with $T_d=20\,$K to the remaining $\sim$ 8000 sources (\citetalias{Dunne2011}). As such, the resulting dust masses may be over or underestimated depending on the `true' dust temperature for each source. Although many studies of nearby galaxies have found that 20\,K is a good approximation
of the dust temperature for most galaxies (e.g. \citealt{Dunne2001,Draine2007,Bendo2010,Boselli2010}). \cite{Clark2015} found that for a sample of local FIR selected galaxies, the average dust temperature was significantly lower at 14.6\,K.

\begin{figure}
\begin{center}
 \includegraphics[trim=0mm 0mm 0mm 0mm clip=true,width=0.98\columnwidth]{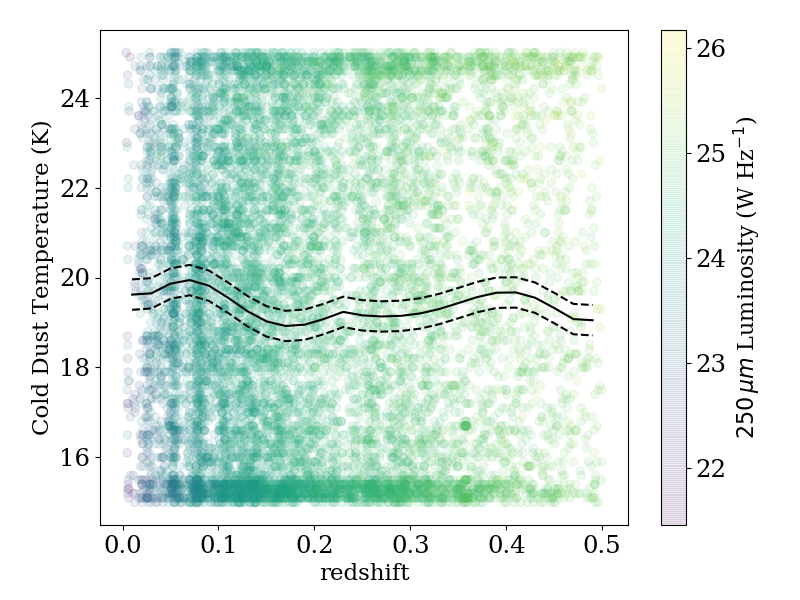}
 \includegraphics[trim=0mm 0mm 0mm 0mm clip=true,width=0.98\columnwidth]{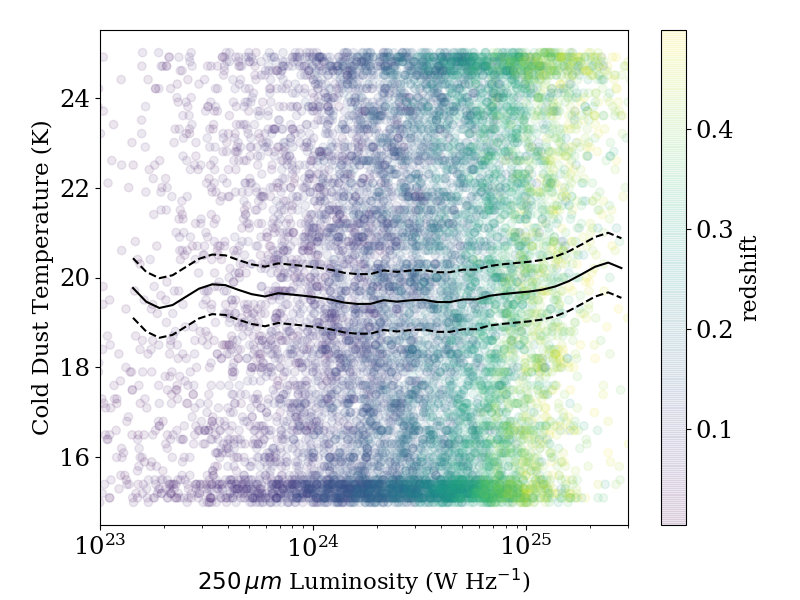}
 \caption{{\it Top:} The cold dust temperature for the sample of galaxies returned by \magphys~\citep{Eales2018} as a function of redshift, coloured by their 250$\,$\micron~luminosity. The mean dust temperature as a function of $z$ is shown as a black line with dashed lines to indicate the 1\,$\sigma$ error on the mean. {\it Bottom:} The cold dust temperature versus 250$\,$\micron~luminosity coloured by redshift. Here we only show those galaxies which have a detection in one of the PACS bands or at 350\,\micron~as well as the 250\,\micron~detection.}
 \label{fig:evo_mphys_z_L}
 \end{center}
\end{figure}

The relationship derived here may not be valid over the redshift range probed in this work. Table~\ref{tab:redshifts} shows that the fraction of galaxies with an individual dust mass measurement from \magphys~decreases rapidly after a redshift of 0.3 and we become increasingly reliant on the $L_{250}-M_d$ relation derived from our training set beyond this point. To check the effect of this, we split the dataset into five redshift slices and re-fitted $L_{250}-M_d$. We found a negligible change in the resulting value of $C$ (<0.01) suggesting that the average dust temperature as a function of redshift is stable in this range. In Figure \ref{fig:evo_mphys_z_L} we show the evolution of the best-fitting cold dust temperature from \magphys~with redshift (top) and $L_{250}$ (bottom) for those galaxies which have either a 4\,$\sigma$ detection in either a PACS band, or at 350\,\micron~as well as a 4\,$\sigma$ detection in $L_{250}$. We do not see any evolution of cold dust temperature with either redshift or 250\,\micron~luminosity. The cluster of galaxies seen at the coldest dust temperatures is likely due to \magphys\ returning the \emph{lower limit of its cold dust temperature prior} (which is set to be flat across the range 15 -- 25\,K). This accounts for approx. 18\,per\,cent of the sample. We will return to this later.

\subsection{A Stacking Analysis of the H-ATLAS Equatorial Field Sample}
\label{sec:stacking}
A high portion of galaxies in our sample have only one \herschel~band with a $>4\sigma$ measurement. Fitting a modified blackbody to these individual SEDs might therefore produce biased dust mass measurements. For this reason we sought a different way to quantify the dust properties of galaxies, and so we perform a stacking analysis on the \herschel~luminosities to derive dust properties in different 250$\,$\micron~luminosity and redshift ($L-z$) bins.

\subsubsection{Calculating Stacked Luminosities}
\label{sec:calc_stacked_L}

We initially calculate 250\,\micron~luminosities for each galaxy assuming a general modified blackbody SED shape with $\beta=2$ and $T_d=20\,\rm K$. We then split the galaxies into 5 equal redshift bins in increments of 0.1. From there we split the lowest redshift bin into 5 equal size log\,$(L)$ bins. To ensure that there are sufficient number statistics in any given bin, if any log\,$(L)$ bin has fewer than 20 galaxies it is aggregated into the next closest log\,$(L)$ bin. For the remaining redshift bins we put all of the galaxies into one log\,$(L)$ bin. Binning in this way gives nine $L-z$ bins. We initially used two even log$(L)$ bins for these redshift slices; however, we found that this resulted in having one very densely populated bin and one very sparsely populated one. Grouping the most extreme galaxies in each redshift slice meant that the masses and temperatures derived for these galaxies were noisy. Splitting into the nine $L_z$ bins means that we can probe any trends in dust temperature with redshift or luminosity, whilst ensuring that the SED shapes of the galaxies in any bin are likely to be similar ie the stacked SED ought to be representative of most of the galaxies in the $L-z$ bin.

Traditionally for this kind of analysis, the inverse variance weighted average (IVWA, \citealt{hartung2011statistical}) would be used to find the stacked luminosities in each bin since it gives the lowest variance estimate of the average. The inverse variance weighted average ($\hat{y}$) of a quantity $y$ is given by:

\begin{equation}
\hat{y} = \frac{\sum_N y_i/\sigma_i^2}{\sum_N 1/\sigma_i^2},
\end{equation}

\noindent where $N$ is the number of independent observations, and $y_i$ and $\sigma_i$ are the value and variance of the $i^{\rm th}$ measurement. The variance ($D^2(\hat{y})$) associated with the IVWA is given by:

\begin{equation}
D^2(\hat{y}) = \frac{1}{\sum_N 1/\sigma_i^2}.
\end{equation}

Comparing the median and mean estimates of the average, it became apparent that whilst the IVWA may give the answer with the least variance, this estimate is biased when used for values with a large (orders of magnitude). In the case where a constant uncertainty is assumed, galaxies with a lower signal, and therefore lower absolute noise, will be erroneously up-weighted using this technique. We therefore opt to use the median, which is not biased in this way.

At this step, in order to minimise the effect of using a poor representation of the SED shape to find the $k$-correction, we simply apply the correction to shift the SED of the stacked bin to the median redshift in each bin rather than to $z=0$, this is a smaller correction and so will introduce less bias. We account for this redshift by adjusting the temperatures by $(1+z)$ within the modified blackbody fitting stage. Once the SED has been fitted with a modified blackbody, we re-calculate the luminosities of each $L-z$ bin assuming that the shape is well-represented by the best-fitting modified blackbody.

The statistical uncertainties for the stacked luminosities with more than 500 galaxies in each bin are estimated using the \cite{Gott2001} method.  In brief, given $N$ measurements $M_{i}$ in order of value, then the probability that the median of the underlying population from which the sample is drawn lies between $M_{i}$ and $M_{i+1}$ is

\begin{equation}
P = \frac{2^{N} N!}{i!(N-i)!}.
\end{equation}

\noindent We then define $r=i/N$, and $M(r)=M_i$. The expectation value of $r$ is simply 0.5, and the standard deviation is given by $1/(4N)^{0.5}$. For bins with fewer galaxies we perform a simple bootstrapping analysis where we resample with replacement and use the bootstrap error as described in \citetalias{Beeston2018}. To this statistical uncertainty, we add the calibration errors for PACS and SPIRE in quadrature, which are 7\% and 5.5\%, respectively.

\subsubsection{Fitting Modified Blackbodies to the Stacked SEDs}

To find dust properties for the galaxies in each $L-z$ bin we attempt to fit both one and two component modified blackbodies to the stacked SEDs. To find the best-fitting SED shapes we use the python package `lmfit', specifically its implementation of the Monte Carlo Markov Chain (MCMC) python package `emcee' \citep{lmfit,emcee}. The posterior distributions are sampled by emcee, and the user sets up a log-posterior probability, essentially calculating the probability that the combination of parameters at the current step is the represents the `true' values, given the observed data.  We use the median of the probability distributions for each parameter as the best-fitting value, and the 16$^{\rm th}$ and 84$^{\rm th}$ percentiles for the 1$\,\sigma$ uncertainty estimates. The equations describing the modified blackbody (MBB) functions are:

\begin{equation}
L_{\nu} = 4 \pi \kappa_{\nu}(\beta) M_d B(\nu, T_d),
\label{eq:simple_lum}
\end{equation}

\noindent for a single component MBB, where the dust emissivity spectral index ($\beta$), dust mass ($ M_d$) and dust temperature ($T_d$) are allowed to vary, and:

\begin{equation}
L_{\nu} = 4 \pi \kappa_{\nu}(\beta) \left[ M_{\rm d, w} B(\nu, T_{\rm d, w}) + M_{\rm d, c} B(\nu, T_{\rm d, c}) \right]
\end{equation}

\noindent for a two component MBB, where $T_{\rm d, w}$ and $T_{\rm d, c}$ are the warm and cold temperatures respectively, and $M_{\rm d, w}$ and $M_{\rm d, c}$ are the warm and cold mass components respectively. We choose to limit our dust temperatures to the same values assumed by \magphys\ for the cold dust component ($15-25\rm\,K$) but assume a wider range of $20-60\rm\,K$ for the warm dust temperature instead of the $30-60\rm\,K$ used by \magphys)  The warm dust temperature of individual galaxies can only be constrained when PACS data are present, however, most individual galaxies are not detected by PACS in our sample, particularly at redshifts $>0.2$ where the fraction drops to $<5$\,per\,cent. The stacking analysis is therefore necessary to probe the evolutionary trends in our data since so few galaxies at higher redshifts will have a sufficient signal-to-noise in either of the PACS bands to constrain the warm component through fitting the SEDs of individual galaxies.

Our main concern is not how much mass is assigned to each component but rather the total mass, we therefore instead choose to fit the total mass and the fraction of mass in the cold component by sampling the total mass in the MCMC rather than the warm and cold masses. We also marginalise over the warm temperature by finding the optimum mass-weighted temperature ($T_{\rm d, MW}$). This is defined as the average of the cold and warm components as weighted by their individual masses, and is given by:

\begin{equation}
T_{\rm d, MW} = \frac{M_{\rm d, c} T_{\rm d, c} + M_{\rm d, w} T_{\rm d, w}}{M_{\rm d, c} + M_{\rm d, w}} = T_{\rm d, c} + 
\left(T_{\rm d, w} - T_{\rm d, c}\right) \left(1 + \frac{M_{\rm d, c}}{M_{\rm d, w}} \right)^{-1}.
\end{equation}

Marginalising the parameters in this way gives two advantages: firstly, the warm component properties will always be much noisier than the cold component. This is because the constraints on the warm dust component are weaker than the cold dust component since its emission will peak at higher frequencies where we have less coverage. Secondly, the total mass and mass-weighted temperature are better indicators of the overall dust properties corresponding to a two component MBB than individual warm and cold components.

\begin{figure}
 \includegraphics[trim=10mm 5mm 5mm 0mm clip=true,width=\columnwidth]{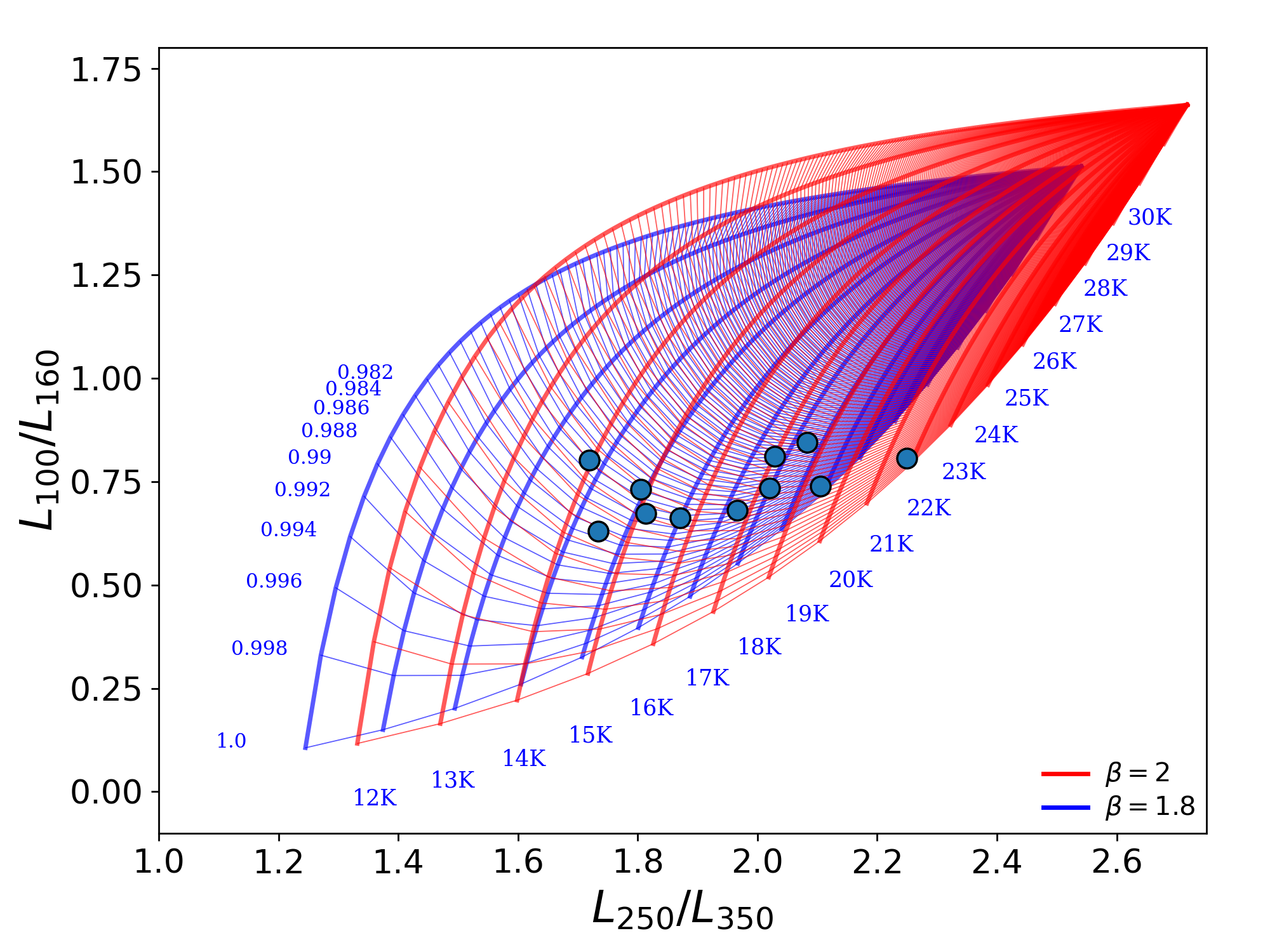}
 \caption{The observed 100$\,$\micron/160$\,$\micron~colours for the $H$-ATLAS sample stacked SEDs in the $L-z$ bins described in Section \ref{sec:calc_stacked_L} against their observed 250$\,$\micron/350$\,$\micron~colours. We show grids of evenly increasing cold temperature (text at bottom of the grid) and cold/warm mass ratio (text at the top of the grid) with a constant warm temperature of 40$\,$K for $\beta$ values of 1.8 (cyan), and 2 (red) as though observed at $z=0.5$.}
 \label{fig:colour_grid}
\end{figure}

We allow $\beta$ to vary for the one component MBB since it can improve the fit to the SED shape. However, the data is not sufficient to constrain $\beta$ for the two component MBB. Instead we simulate grids of FIR colours (250$\,$\micron/350$\,$\micron~and 100$\,$\micron/160$\,$\micron) assuming a fixed $\beta=1.8$ and $\beta=2$ with fixed $T_{\rm d, w} = 40\,\rm K$, $T_{\rm d, c}$ varying between 12 and 30$\, \rm K$, and a cold to warm mass ratio varying between 0 and 1. The sets of simulated grids are shown in Figure \ref{fig:colour_grid}. The observed colour ratios of the stacked 250$\,$\micron/350$\,$\micron~and 100$\,$\micron/160$\,$\micron~corrected to the bin centre are also shown for comparison. (Here we show the grids as they would appear if redshifted to 0.5; we choose to redshift the grid rather than show rest-frame colours since this does not rely on $k$-corrections based on SED fits.) We see that the simulated grid with fixed $\beta=2$ is a better fit to the parameter space sampled by the stacked observations, and so we choose to use this value in the two temperature component stacked fits. When deriving dust masses for the stacked bins, we use the \cite{James2002} dust mass absorption coefficient $\kappa_{850\,\mu \rm m} =0.077 \rm m^{2}\,kg^{-1}$ in line with that used in \citetalias{Beeston2018}, \citetalias{Dunne2011} and \citet{Driver2018}. 
Hereafter we will refer to the one and two temperature MBB fits as 1MBB and 2MBB respectively.  

\begin{table*}
\centering
    \begin{tabular}{ccc|ccccccc} \hline
$\Delta z$ & $\Delta$log\,$L_{250}$  & $N$ & $T_{\rm d, W}$ & $T_{\rm d, C}$  & $T_{\rm d, MW}$ & log\,$M_{\rm w}$  & log\,$M_{\rm c}$ & log\,$M_{\rm d, tot}$ & $\chi^2$ \\ 
           &  ($\rm W\,Hz^{-1}$)  &      &    (K)         &       (K)         &         (K)        &   ($M_{\odot}$)  &   ($M_{\odot}$)   &  ($M_{\odot}$)  & \\ \hline 
0.002 - 0.1 & 21.6 - 22.2 & 27 & 27.5$\pm$8.2 & 15.2$\pm$0.6 & 16.3$\pm$0.5 & 4.44$\pm$0.04 & 5.45$\pm$0.02 & 5.49$\pm$0.04 & 1.29 \\
0.002 - 0.1 & 22.2 - 22.9 & 167 & 25.2$\pm$2.4 & 15.1$\pm$0.2 & 16.2$\pm$0.2 & 5.56$\pm$0.02 & 6.48$\pm$0.01 & 6.53$\pm$0.01 & 2.92 \\
0.002 - 0.1 & 22.9 - 23.6 & 1147 & 25.1$\pm$3.5 & 16.0$\pm$0.6 & 17.7$\pm$0.4 & 6.44$\pm$0.03 & 7.09$\pm$0.01 & 7.18$\pm$0.03 & 1.18 \\
0.002 - 0.1 & 23.6 - 24.3 & 2201 & 25.4$\pm$4.2 & 19.0$\pm$1.0 & 20.5$\pm$0.5 & 6.83$\pm$0.06 & 7.36$\pm$0.05 & 7.48$\pm$0.02 & 0.58 \\
0.002 - 0.1 & 24.3 - 25.0 & 199 & 25.6$\pm$4.7 & 21.9$\pm$1.6 & 22.9$\pm$0.5 & 7.34$\pm$0.14 & 7.75$\pm$0.14 & 7.89$\pm$0.02 & 0.57 \\
0.1 - 0.2 & 23.7 - 25.3 & 7739 & 29.7$\pm$19.4 & 19.9$\pm$1.6 & 20.7$\pm$0.5 & 6.88$\pm$0.08 & 7.93$\pm$0.07 & 7.97$\pm$0.02 & 0.18 \\
0.2 - 0.3 & 24.2 - 25.4 & 6365 & 49.0$\pm$77.2 & 20.9$\pm$0.4 & 21.1$\pm$0.3 & 6.29$\pm$0.02 & 8.49$\pm$0.0 & 8.49$\pm$0.02 & 0.26 \\
0.3 - 0.4 & 24.5 - 26.1 & 5861 & 57.1$\pm$75.2 & 21.1$\pm$0.2 & 21.3$\pm$0.2 & 6.46$\pm$0.01 & 8.86$\pm$0.0 & 8.86$\pm$0.01 & 0.77 \\
0.4 - 0.5 & 24.7 - 25.8 & 5535 & 58.1$\pm$73.7 & 21.3$\pm$0.2 & 21.5$\pm$0.2 & 6.74$\pm$0.01 & 9.18$\pm$0.0 & 9.18$\pm$0.01 & 0.65 \\ \hline \hline
\end{tabular}
\caption{The best-fitting SED parameters derived for the stacked galaxies in each of our $L-z$ bins using the two temperature component modified blackbody fits. The value quoted is the median $\pm$ 16th and 84th confidence intervals. The equivalent information for the one component MBB fit is given in Table~\ref{tab:stacked_SED_params_one}.}
\label{tab:stacked_SED_params}
\end{table*}

\subsubsection{Results from the Stacked SEDs}
The best-fitting SED parameters for all $L-z$ bins for the two component MBBs are listed in Table \ref{tab:stacked_SED_params}. The differences in the one and two temperature component MBB fits are discussed further in Appendix~\ref{app:fits} with examples of the best-fits in Figures~\ref{fig:fitgrid1} and \ref{fig:corner}, and the results for the one component MBB fit provided in Table~\ref{tab:stacked_SED_params_one}.  
\begin{figure*}
\includegraphics[trim=20mm 0mm -20mm 0mm clip=true,width=1.0\columnwidth]{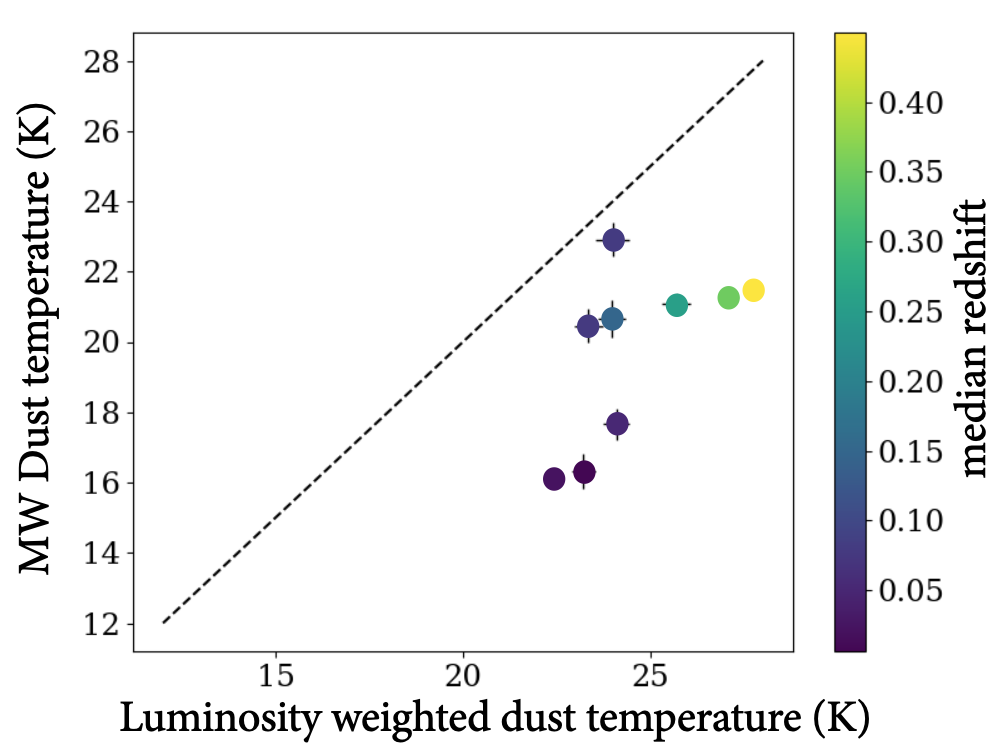}
\includegraphics[trim=10mm 0mm 0mm 0mm clip=true,width=0.95\columnwidth]{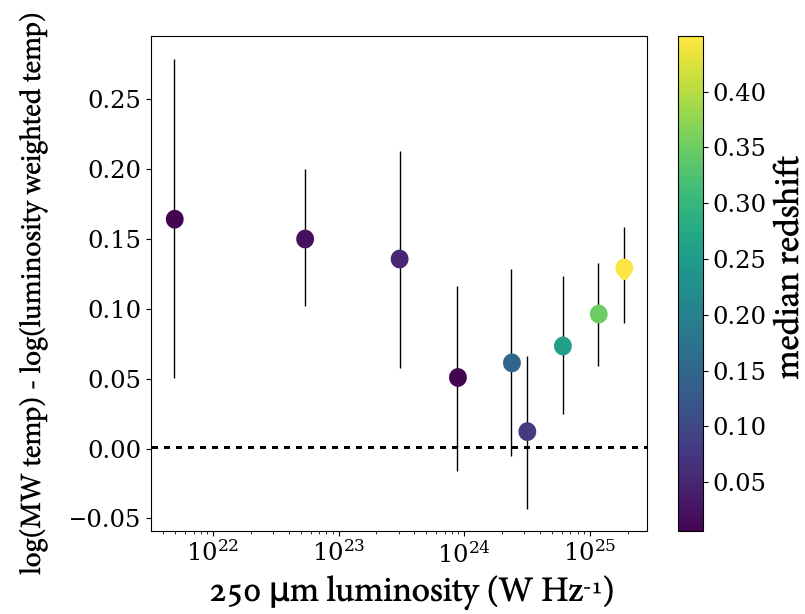}
 \caption{{\it Left:} A comparison of the temperature from the one and two temperature component (luminosity-weighted and mass-weighted respectively) MBB fits to the stacked SEDs. The dashed line shows the one-to-one relationship. {\it Right:} The difference between the temperature from the two component (mass-weighted) and the one component MBB fits (luminosity-weighted) with 250\,$\mu \rm m$ luminosity. The data points are coloured by the median redshift in that bin.   }
 \label{fig:onevtwo}
\end{figure*}

The dust temperature from the one component MBB fit (1MBB) is essentially luminosity-weighted, since it accounts for the majority of the luminosity, unlike the mass-weighted temperature from the 2MBB which accounts for most of the mass. Since $k$-correcting an SED with one temperature component is simpler than one with two components, we choose to calculate the rest-frame $L_{250}$ of each galaxy using the $k$-corrections based on the 1MBB fit of its $L-z$ bin. The 1MBB gives a good representation of the shape of the stacked SEDs and so using this luminosity-weighted estimate of the temperature will reproduce the shape of the SED much more reliably than the mass-weighted temperature. In order to calculate dust masses for each galaxy, we use its rest-frame 250$\,$\micron~luminosity and the mass-weighted temperature of its $L-z$ bin along with Equation \ref{eq:simple_lum}. Hereafter we will refer to the dust temperatures from the 1MBB and 2MBB fits as luminosity-weighted and mass-weighted temperatures respectively.

Figure \ref{fig:onevtwo} (left) shows that the dust temperatures derived from the 1MBB fits can be vastly different to the mass-weighted temperatures derived from the 2MBB fits for each $L-z$ bin (see example SEDs in  Figure~\ref{fig:fitgrid1}). The largest offsets from the one-to-one relationship are seen in the SEDs with the lowest mass-weighted dust temperatures (these $L-z$ bins are the three lowest 250-$\mu$m luminosity bins in the lowest redshift slice), where there is evidence for dust temperatures at $<$20\,K. These stacked SEDs are similar to those dust-rich (in comparison to their stellar mass) sources with dust temperatures 13$-$20\,K found in the 250\,\micron~selected blind survey of \cite{Clark2015}. We are likely only sensitive to these sources in the lowest redshift slice since the galaxies contained in these $L-z$ bins drop out of our survey due to having low absolute dust masses ($<10^{6}\,M_{\odot}$) and cold dust temperatures (therefore lower 250\,\micron~luminosities). This discrepancy in $T_d$ (one component MBB fit) and $T_{\rm d, MW}$ (two component MBB fit) for these low luminosity and cold dust temperatures results in an offset of $\sim 0.16\,$dex in dust mass between the fits to the stacked data as shown in Figure \ref{fig:onevtwo} (right). At higher luminosities, the offset is smaller but still positive.

\subsubsection{Dust Temperature with Redshift}
\begin{figure*}
 \includegraphics[trim=10mm 0mm 0mm 0mm clip=true,width=0.96\columnwidth]{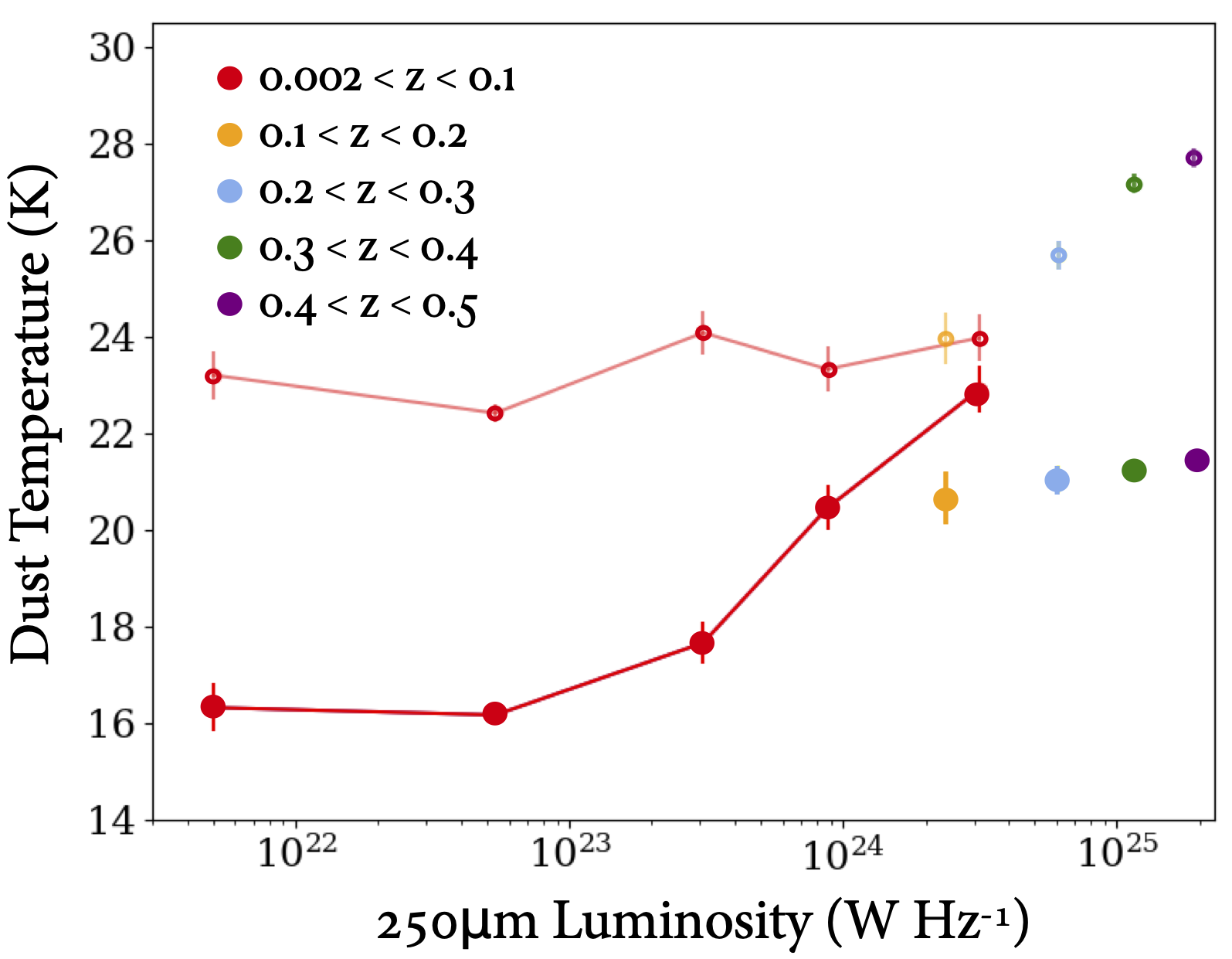}
  \includegraphics[trim=0mm 0mm 10mm 0mm clip=true,width=0.96\columnwidth]{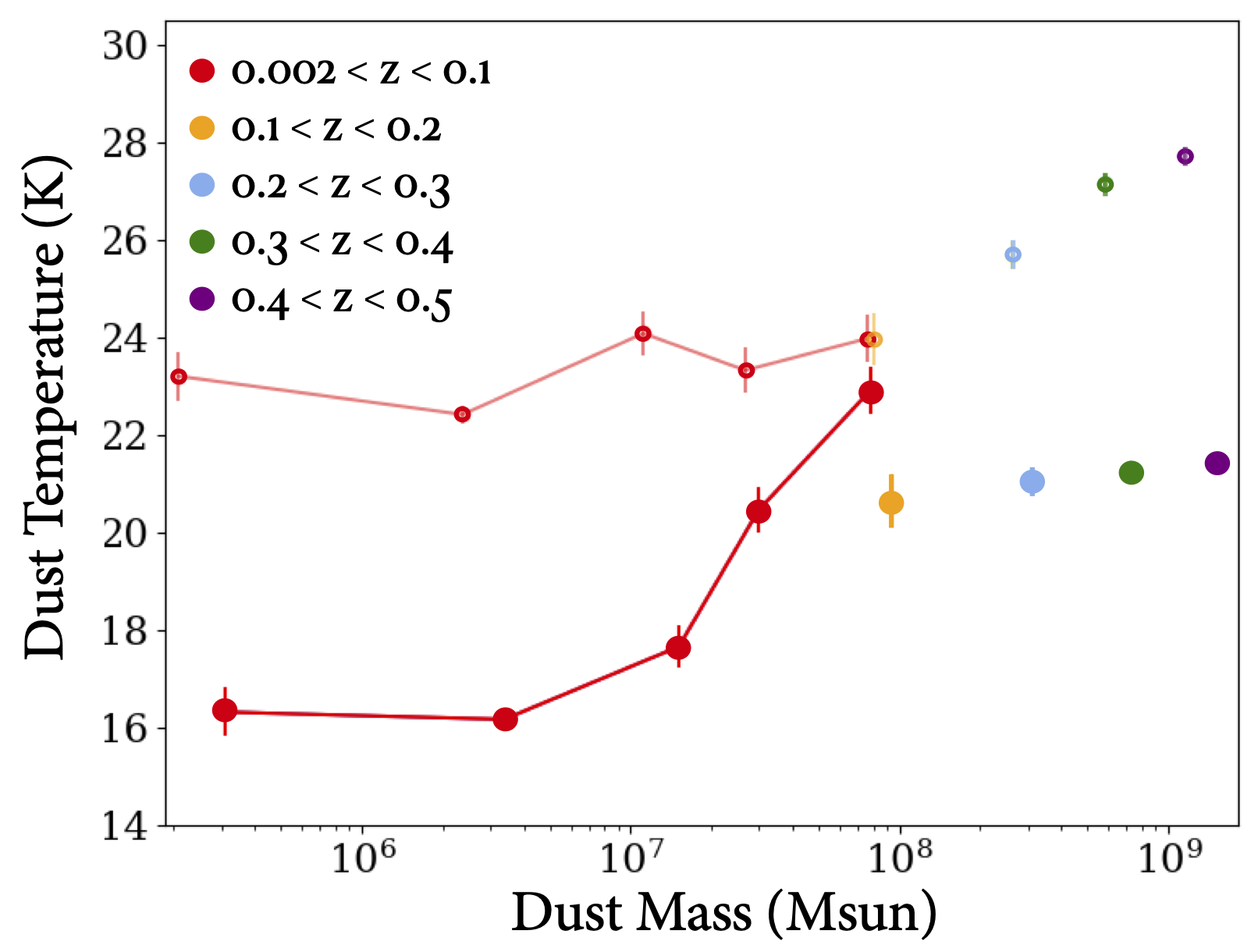}
 \caption{The mass-weighted (large opaque points) and luminosity-weighted (smaller transparent points) dust temperatures from the best-fitting stacked SEDs as a function of {\it left:} 250$\,$\micron~luminosity and {\it right:} the best-fitting dust mass. In each figure the temperature value comes from the median of the probability distribution for each $L-z$ bin, and the uncertainties come from the 16$^{\rm th}$ and 84$^{\rm th}$ percentile values from these probability distributions.  For the lowest redshift slice $0.002< z<0.1$ (where more than one $L-z$ bin is used), we link the points with a straight line to emphasize the evolution of temperature in this slice.   The solid translucent points represent the mass-weighted temperatures, and the smaller empty datapoints represent the luminosity-weighted temperature. }
 \label{fig:TdMW_evo_Md_L}
\end{figure*}

Both \cite{Dunne2000} and \cite{Dale2001} reported that the dust temperatures of nearby galaxies ($z<0.1$) seem to have a strong correlation with their IR luminosity. The same correlation is not seen at higher redshifts (see e.g. \citealt{Coppin2008,Symeonidis2009,Amblard2010,Rex2010,Seymour2010,Symeonidis2011}) though the temperature dependence observed in the evolution of the IR luminosity function at high $z$ is attributed to `cold SED' galaxies\footnote{A`cold SED' in this scenario is defined as galaxies where the rest-frame wavelength of the IR peak is longer than 90\,$\mu$m.} changing more rapidly with redshift in comparison to warmer galaxies.

Is there evidence for any evolution in the dust temperature in the FIR selected galaxies in this work? In Figure \ref{fig:TdMW_evo_Md_L} we compare the mass-weighted and luminosity-weighted temperatures with $L_{250}$ and $M_d$ in different redshift bins. The mass-weighted temperatures in the lowest redshift slice are a strong function of $L_{250}$ and $M_d$. This is unsurprising since dust becomes increasingly bright per unit mass with increasing temperature, and where there is more dust we will naturally expect more emission. We see a shallow and steady increase in mass-weighted temperature for each subsequent redshift slice, but since we also see an increase in $L_{250}$ in these bins we cannot disentangle whether redshift or luminosity is the most important factor in this increase. 

Unlike the results from the \magphys~fits to individual galaxies, our stacking analysis supports an increasing mass-weighted dust temperature with redshift or luminosity.

\begin{figure*}
 \includegraphics[trim=0mm 0mm 0mm 0mm clip=true,width=\columnwidth]{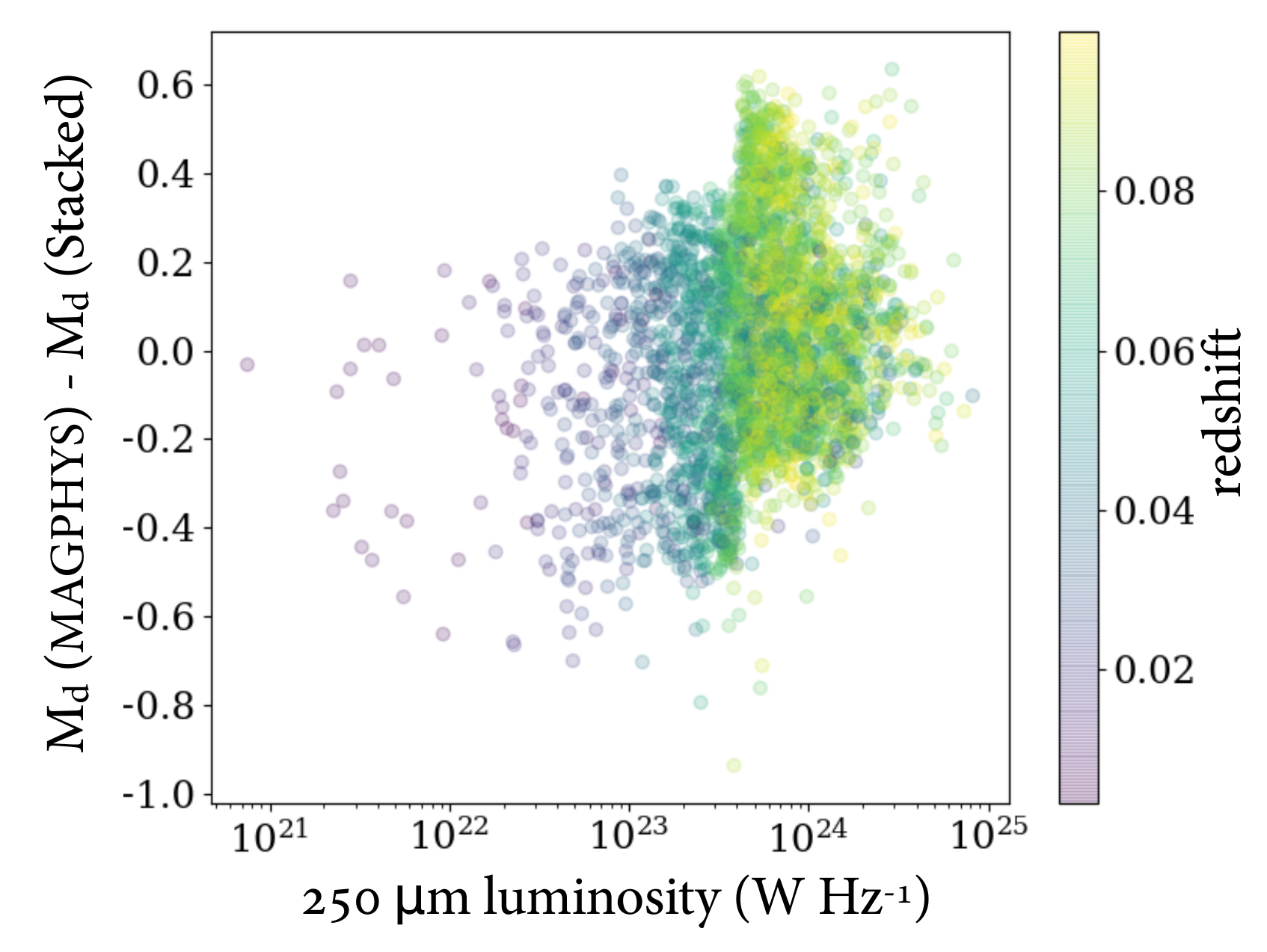}
  \includegraphics[trim=0mm 0mm 0mm 0mm clip=true,width=\columnwidth]{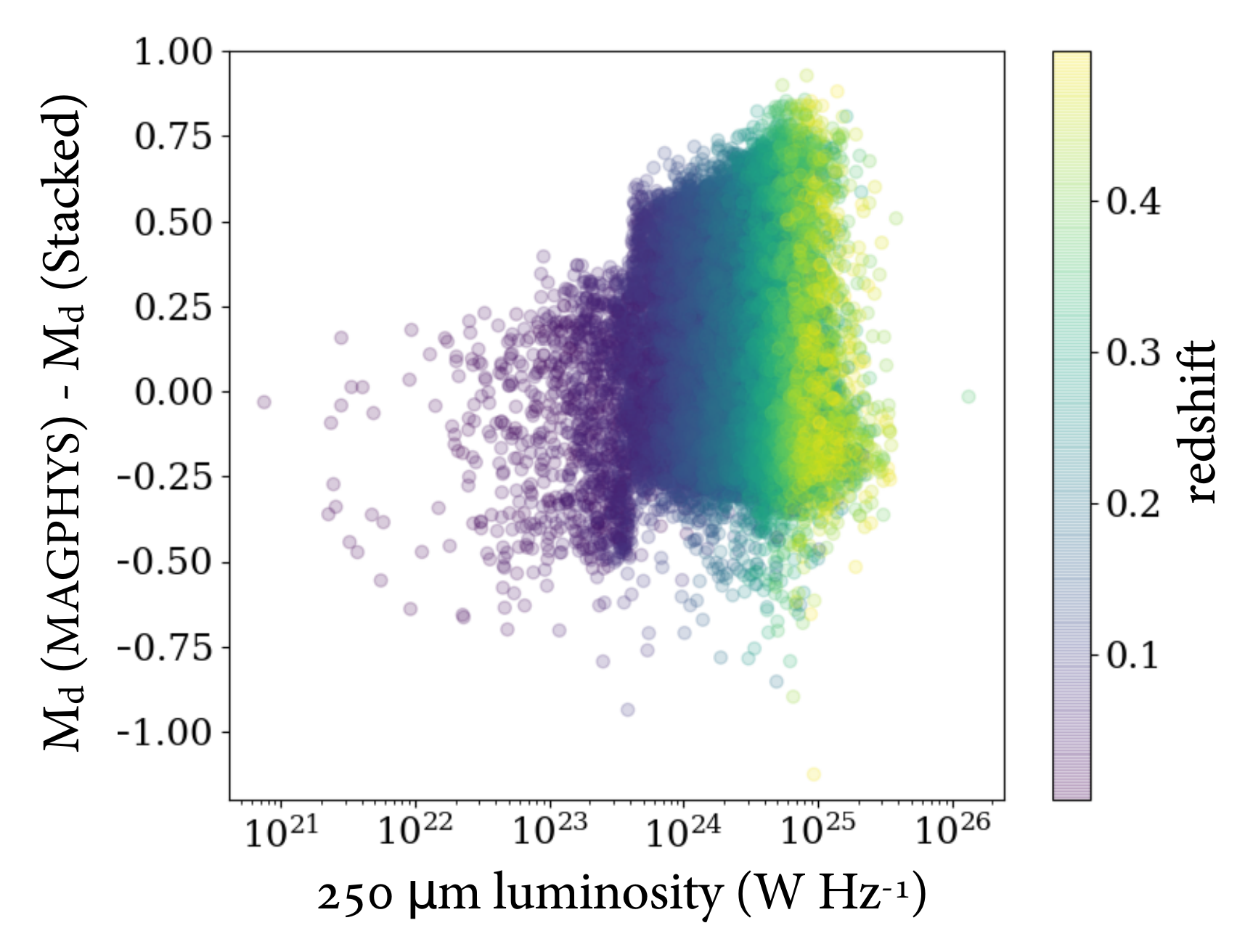}
 \caption{The difference between the \magphys~based {\it log} dust masses and the 2MBB stacked {\it log} dust masses as a function of 250\,\micron~luminosity for ({\it left:}) $z<0.1$ and ({\it right:}) $z<0.5$ (see colourbar).  }
 \label{fig:stacked_magphys_scatter}
\end{figure*}

\begin{figure}
 \includegraphics[trim=0mm 15mm 0mm 0mm clip=true,width=\columnwidth]{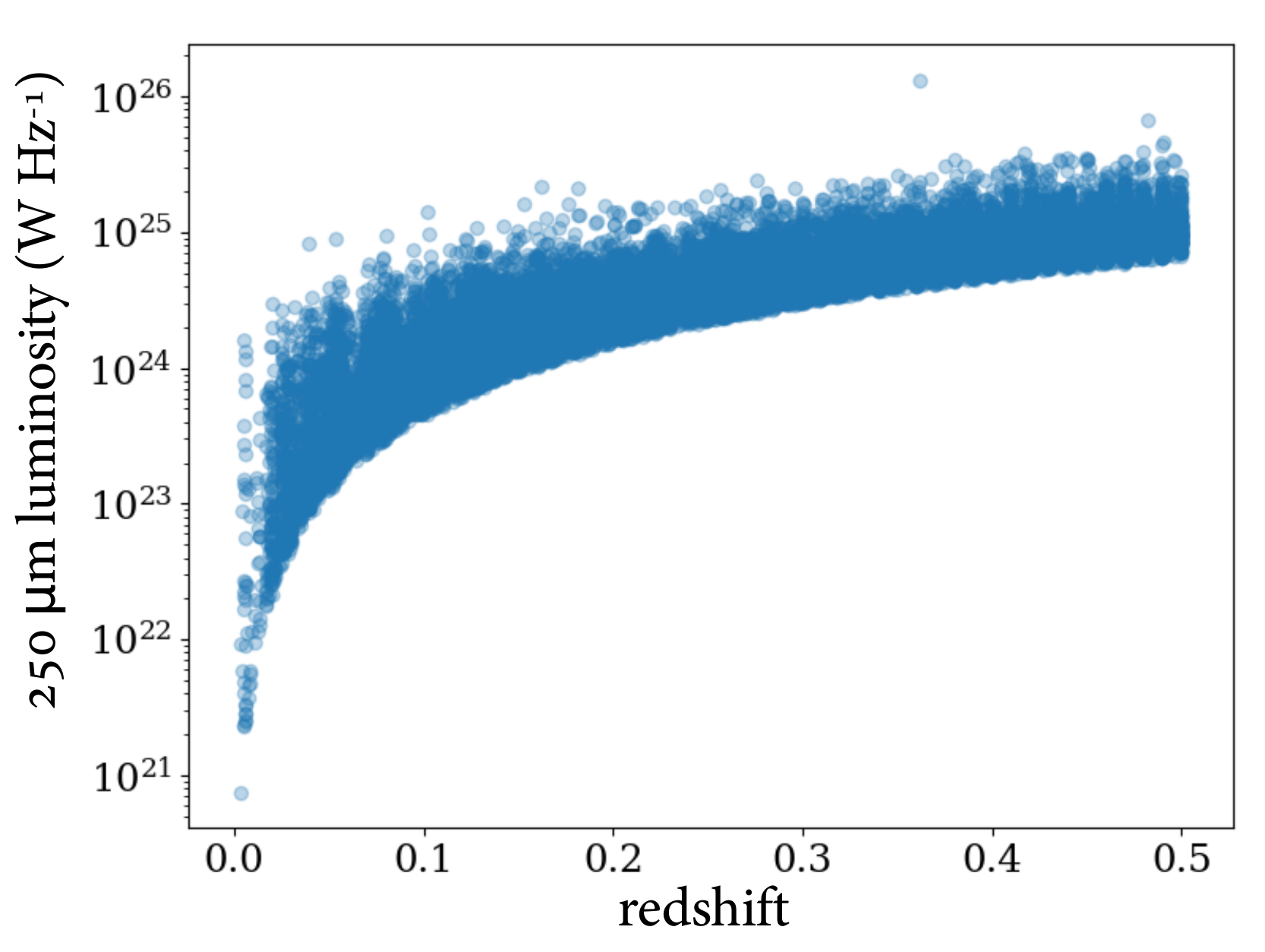}
 \caption{The 250\,\micron~luminosities for the $H$-ATLAS sample as a function of redshift.}
 \label{fig:l250vsz}
\end{figure}

\subsection{Comparison of the Dust Properties from \magphys~and Stacking}

In this section we briefly compare the dust mass estimates for the sample from (i) fitting \magphys~to individual galaxies and (ii) using the 250\,$\mu$m flux density and the stacked SEDs in $L-z$ bins (Figure~\ref{fig:stacked_magphys_scatter}). Generally the dust masses are similar for galaxies in the low-redshift ($z<0.1$) slice with a median offset of only $\sim$0.01\,dex, but with a significant scatter around 1\,dex. Figure~\ref{fig:stacked_magphys_scatter} (left) shows that \magphys~tends to assign lower dust masses at low luminosities and higher masses at high luminosities than the 2MBB stacked dust masses. At higher redshifts (Figure~\ref{fig:stacked_magphys_scatter}, right) the \magphys~dust masses are generally higher. This may point to a bias within either the \magphys~fitting routine, or with the stacking analysis, or even both. In \citetalias{Beeston2018} we briefly discussed the potential flaws of using \magphys~to fit the SEDs of galaxies with faint FIR emission, namely its propensity to return the priors on the parameters. It is probably this effect of returning the median of the prior for the cold temperature (15\,K) that causes the difference in the dust masses, since the mass-weighted temperatures in the higher redshift slices are all above this value. Therefore this suggests that the \magphys\ results will be biased to higher masses. As the cold dust emission is fainter in comparison to warmer emission at higher redshifts, the fits to the high redshift stacked SEDs will be more sensitive to warm dust and less sensitive to cold dust.

In this Section, we derived dust properties of 29,241 FIR-selected galaxies at $z<0.5$ using two different approaches: SED fitting of individual galaxies with \magphys~and SED-fitting of galaxies stacked in bins of $L-z$. There is a tendency for \magphys~to assign lower dust masses to galaxies with low 250-$\mu$m luminosity and higher masses at high $L_{\rm 250}$ in comparison to the two component fits to the stacked SEDs. We suggest this might be due to \magphys~returning the prior dust temperature for sources with faint FIR emission and low signal-to-noise. This also provides an explanation for the lack of evolution in the dust temperature seen when using \magphys~fits (in \citetalias{Dunne2011} and Section~\ref{sec:HATLAS_magphys}) compared to the evolution in dust temperature clearly seen with the stacked method (Figure~\ref{fig:TdMW_evo_Md_L}). 

In Figure~\ref{fig:l250vsz} we show $L_{250}$ as a function of redshift for our full sample in order to illustrate the difficulty of choosing a luminosity range over which to test the evolution of dust properties with redshift. We cannot choose a luminosity slice which is well-populated across all redshifts. We saw earlier that the luminosity-weighted temperature has a shallow evolution with luminosity in the lowest redshift slice and an increase with dust mass, luminosity and redshift in the higher redshift bins.  In the case of both luminosity- and mass-weighted temperature we expect decreased sensitivity to cold dust with redshift since the effective range of rest-frame frequencies will be higher, but it is unlikely that this effect could account for all of the evolution we see. It is not possible to distinguish whether the evolution in dust temperature we see in the stacked method  is driven by increasing redshift or luminosity.

Next we use the dust properties for our galaxies to produce statistical functions of the luminosity and dust mass and search for trends in redshift. 

\section{Deriving Statistical Functions}

\subsection{Completeness Corrections}
\label{sec:completeness}

To derive mass functions for our FIR selected sample we first need to consider the completeness of our survey. Here we outline our method for correcting the number counts of sources using estimates of completeness for the submm and optical catalogues, as well as incompleteness introduced by matching the submm sources to optical counterparts.

\subsubsection{Submillimetre Catalogue Completeness}
The submm catalogue completeness correction ($c_s$) is set by both the flux limit of the survey and the source extraction process used to compile the $H-$ATLAS catalogue. To estimate this, \citet{Valiante2016} simulated sources, added them to the $H-$ATLAS maps and performed the same source extraction technique as used to create the observed source catalogue. This allows one to determine the likelihood that sources are lost to noise. The values for the submm catalogue completeness corrections as a function of 250$\,$\micron~flux are listed in Table \ref{tab:comp}.

\begin{table}
\centering
\begin{tabular}{cccc}
 \hline
$S_{250\micron} (\rm Jy)$ & $c_s$ & $N$ & \% \\  \hline
20.6 - 25.4 & 1.357 & 383 & 1.3 \\
25.4 - 31.2 & 1.151 & 4461 & 15.3 \\
31.2 - 38.3 & 1.073 & 8536 & 29.2 \\
38.3 - 47.0 & 1.029 & 5776 & 19.8 \\
47.0 - 57.8 & 1.012 & 3819 & 13.1 \\
57.8 - 71.0 & 1.009 & 2324 & 7.9 \\
71.0 - 87.2 & 1.006 & 1408 & 4.8 \\  
\hline
$M_{r}$ (mag) & $c_r$ & $N$ & \% \\  \hline
  21.5 - 21.6 & 1.10 & 131 & 0.45 \\
21.6 - 21.7 & 1.14 & 123 & 0.42 \\
21.7 - 21.8 & 1.21 & 118 & 0.40 \\
21.8 - 21.9 & 1.29 & 100 & 0.34 \\
21.9 - 22.0 & 1.42 & 110 & 0.38 \\
22.0 - 22.1 & 1.62 & 82 & 0.28 \\
22.1 - 22.2 & 1.9 & 96 & 0.33 \\
22.2 - 22.3 & 2.33 & 76 & 0.26 \\
22.3 - 22.4 & 5.88 & 72 & 0.25 \\ \hline
$z$ & $c_z$ & $N$ & \% \\  \hline
  0.0 - 0.1        &   1.095    &  3741 &  13.0  \\
  0.1 - 0.2        &   1.140    & 7739  &  26.9  \\
  0.2 - 0.3        &   1.244    & 6365  &  22.1 \\
  0.3 - 0.4      &   1.385    &  5861 & 19.8  \\
  0.4 - 0.5      &   1.451    &   5535  &   18.2 \\ \hline
\end{tabular}
\caption{
Corrections for incompleteness in the submm ($c_s$) for different 250$\,$\micron~flux ranges ($S_{250\micron}$), optical catalogues ($c_r$) for different magnitude ranges ($M_r$) and in the source identification ($c_z$) in redshift bins. $N$ is the number of sources in each flux bin and the final column shows the percentage of the total source catalogue.
}
\label{tab:comp}
\end{table}

\subsubsection{Optical Catalogue Completeness}

Optical data for the galaxies in our sample is taken from the SDSS catalogue, which has a magnitude limit of 22.4 in the $r$-band. Given that the SDSS catalogue is close to 100\,per\,cent completeness to the magnitude $M_r=21.5$\,mag, \citetalias{Dunne2011} fit a linear slope to the logarithm of the number counts of sources in bins between these magnitudes. This fit was extrapolated to fainter magnitudes and compared to observed number counts in order to find the completeness $c_r$. We use the same corrections calculated in \citetalias{Dunne2011} here, and the corrections as a function of absolute $r$-band magnitude ($M_{r}$) are listed in Table \ref{tab:comp}.

\subsubsection{ID Completeness}
\label{sec:IDcompleteness}
\citet{Bourne2016} obtained optical IDs for the $H-$ATLAS sources using a likelihood ratio technique. The completeness $c_z$ of the $H-$ATLAS catalogues was derived from the number of reliable IDs and the number of sources they estimate will have counterparts which will be visible both in the optical and submm catalogues. The completeness corrections ($c_z$) required as a function of redshift are listed in Table \ref{tab:comp}.

\subsection{The Luminosity Function and Estimators}

We use two methods to calculate the luminosity and mass functions. Firstly, the traditional $V_{\rm max}$ method \citep{Schmidt1968}, and secondly the method proposed by \cite{Page2000} (hereafter the PC00 method) with the important addition of multiplicative corrections for the sources of incompleteness described in Section \ref{sec:completeness}. The $V_{\rm max}$ volume density $\phi$ in $\rm Mpc^{-3}\,dex^{-1}$ is given by:

\begin{equation}
\phi(L_i) = \sum\limits_{n=1}^{N_i} \frac{c_r c_s c_z}{V_{\rm max}},
\label{eq:dmf_sum}
\end{equation}

\noindent where the sum extends over the number of galaxies $N$ in the $i^{\rm th}$ bin, $V_{\rm max}$ is the accessible volume and $c_r$, $c_s$, and $c_z$ are the completeness corrections (Section \ref{sec:completeness}). We calculate the volume accessible to each galaxy using its SED shape, luminosity, and limiting signal-to-noise ratio (SNR). We find the maximum redshift available to each galaxy numerically by minimising the following:

\begin{equation}
\left |{ \frac{L_{\nu} (1+z)}{4 \pi D_{\rm L}^{2} k} - S_{\nu, \rm lim} } \right |
\end{equation}

\noindent where $L_{\nu}$ is the luminosity of the galaxy at frequency $\nu$, $D_{\rm L}$ is the luminosity distance, $k$ is the $k$-correction of the source based on the SED shape of the $L-z$ bin of the galaxy, and $S_{\nu, \rm lim}$ is the limiting flux for which the source would be visible based on the properties of the $H$-ATLAS survey. Although this can be at any wavelength, in practice this tends to the the SNR at 250\,\micron, and so we use four times the uncertainty on the 250\,\micron~flux as our value of $S_{\nu, \rm lim}$. The SED shape properties $T_d$ and $\beta$ were taken from the one-component modified blackbody fits to the stacked SED of appropriate $L-z$ bin for each galaxy.

Next we use a modified version of the PC00 method. Once again we include corrections for the various forms of incompleteness. The PC00 method has the advantage of not needing to use (sometimes poor quality) data to derive the accessible volume, nor does it overestimate the accessible volume for galaxies near the flux limit in each redshift slice \citep{Page2000}.

The PC00 method takes the form:

\begin{equation}
\phi(L_{i}) = \frac{\sum_{n=1}^{N_i} c_s c_z c_r}{\int^{L_{\rm max}}_{L_{\rm min}} \int^{z_{{\rm max} (L)}}_{z_{\rm min}} \frac{{\rm d}V}{{\rm d}z} {\rm d}z \, {\rm d}L}.
\end{equation}

The quantity ${\rm d}V{\rm d}z$ refers to the path the galaxies in the bin take through luminosity-volume space with redshift. $L_{\rm min}$ and $L_{\rm max}$ are the minimum and maximum luminosities of the bin, $z_{\rm min}$ and $z_{{\rm max}}(L)$ are the minimum redshift of the slice, and the maximum redshift to which a source with luminosity $L$ could be detected within the flux limit with a given $k$-correction, but is not allowed to exceed the maximum redshift of the slice. In essence here the volume is taken from the integral under the curve a source with given intrinsic properties would take through the $L-z$ plane, rather than just using the single value corresponding to the exact redshift at which the source happens to lie.

PC00 initially just presented a version of this estimator where all galaxies in a bin would follow the same $L-z$ relationship. \citetalias{Dunne2011} modified this to allow each galaxy to trace a unique path across the $L-z$ plane. This is more realistic since the SED properties of each galaxy can be different, as well as the complication that the $H-$ATLAS selection is based on the signal-to-noise ratio (SNR) of a source rather than a single limiting flux across the catalogue, so a different limiting flux can also be employed. This modified PC00 method takes the form:

\begin{equation}
\phi(L_{i})  = \sum\limits_{n=1}^{N_i}  \frac{c_s c_z c_r}{\int^{L_{\rm max}}_{L_{\rm min}} \int^{z_{{\rm max}, i}}_{z_{\rm min}} \frac{{\rm d}V}{{\rm d}z} {\rm d}z \, {\rm d}L}
\end{equation}

where $z_{{\rm max}, i}$ is the maximum redshift across the luminosity bin for galaxy $i$, as a function of luminosity, limiting flux, and the temperature assumed for the MBB fit to the SED. Allowing the accessible volume to evolve across the luminosity bin is most effective when considering those galaxies which lie close to the boundary at which they would fall out of the survey. 

To derive the luminosity function, we estimate the space density of galaxies in a given luminosity bin by bootstrapping with replacement 1000 times. This produces 1000 values for each luminosity bin of the LF from which we estimate the mean and uncertainty to produce the data points and error bars in Figure \ref{fig:LF_PC00vVmax}.  We then fit a Schechter function (SF) to each LF bootstrap realisation to quantify the statistical uncertainty on the best-fitting parameters. In ${\rm log}L$ space, the SF takes the form:

\begin{align}
  S(L; \alpha, L^{*}, \phi^{*}) & = \phi^{*} e^{-10^{{\rm log}L - {\rm log}L^*}} \nonumber \\
 & \times \left(10^{{\rm log}L - {\rm log} L^*}\right)^{\alpha+1} d\,{\rm log}L,
\end{align}

\noindent where we have explicitly included the factor ${\rm ln} 10$ in the definition of $\phi^{*}$, such that $\phi^{*}$ is in units of $\rm Mpc^{-3}\,dex^{-1}$.  We use the individual best-fitting SF fits to find uncertainty estimates for each SF parameter. We also estimate the error in each redshift bin due to cosmic variance using the estimator from \cite{Driver2010}\footnote{\url{cosmocalc.icrar.org}} for the full survey volume.  Naturally this uncertainty will vary across the different bins due to the different maximum volumes available to the galaxies in each bin. We choose therefore to quote the uncertainty due to cosmic variance separately from the statistical uncertainty.

\subsection{The Dust Mass Function}
\label{sec:dmf}

We estimate the dust mass function (DMF) using both the $V_{\rm max}$ and PC00 methods and the masses from the two methods - \magphys~and stacking - described in Section~\ref{sec:dustprops}. For the \magphys~method, we estimate the space density of galaxies in each bin by bootstrapping the space density in the same way as described for the luminosity function. The error bars on each data point are derived using the bootstrap with additional perturbation from the \magphys~uncertainties in individual galaxy dust masses.

For the stacking method we use the SED shape derived from the two dust temperature component fit (2MBB) to the stacked SED in each $L-z$ bin and the 250\,$\mu$m luminosity for each galaxy to derive its dust mass. Errors are determined using MCMC fits.

\section{Results}

\subsection{The Luminosity Function}

\begin{figure}
  \includegraphics[trim=0mm 0mm 0mm 0mm clip=true,width=\columnwidth]{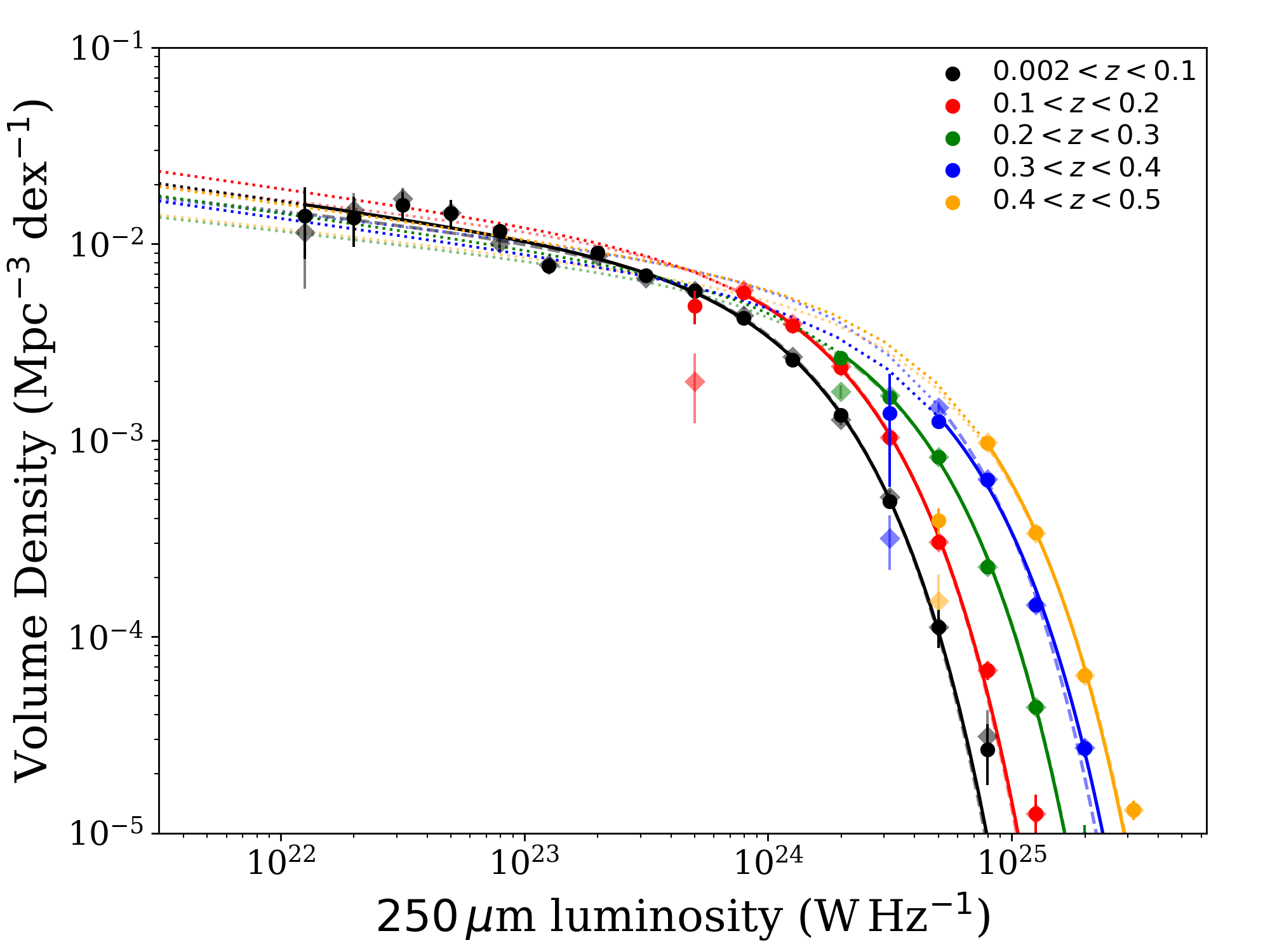}
 \caption{The 250\,\micron~luminosity function (LF) based on stacking assuming $V_{\rm max}$ (transparent diamonds), and PC00 (opaque circles) estimators. The LF in five redshift slices are indicated by the different colours (see legend). Schechter fits to the LFs for each redshift slice are shown as solid (PC00) and dashed ($V_{\rm max}$) curves. Error bars are derived from a bootstrap analysis where the variance of 1000 realisations of each LF determines the uncertainty on each datapoint. }
 \label{fig:LF_PC00vVmax}
\end{figure}

In Figure \ref{fig:LF_PC00vVmax} we show the difference in the resulting luminosity functions derived from the PC00 and $V_{\rm max}$ LFs with redshift. The largest difference is seen in the lowest luminosity bins which produce lower dust masses for the $V_{\rm max}$ LF. This nicely illustrates the bias seen in the $V_{\rm max}$ method compared to the PC00.  The median of the best-fitting SF fit parameters are listed in Table \ref{tab:schechterTabLF}. Since there is not enough
data below the knee of the LF to determine the low mass slope for the higher redshift slices, we keep $\alpha$ constant for redshifts beyond $z= 0.1$ and set it to the value fit to the lowest redshift bin. 

\begin{figure}
 \includegraphics[trim=15mm 0mm 0mm 0mm clip=true,width=\columnwidth]{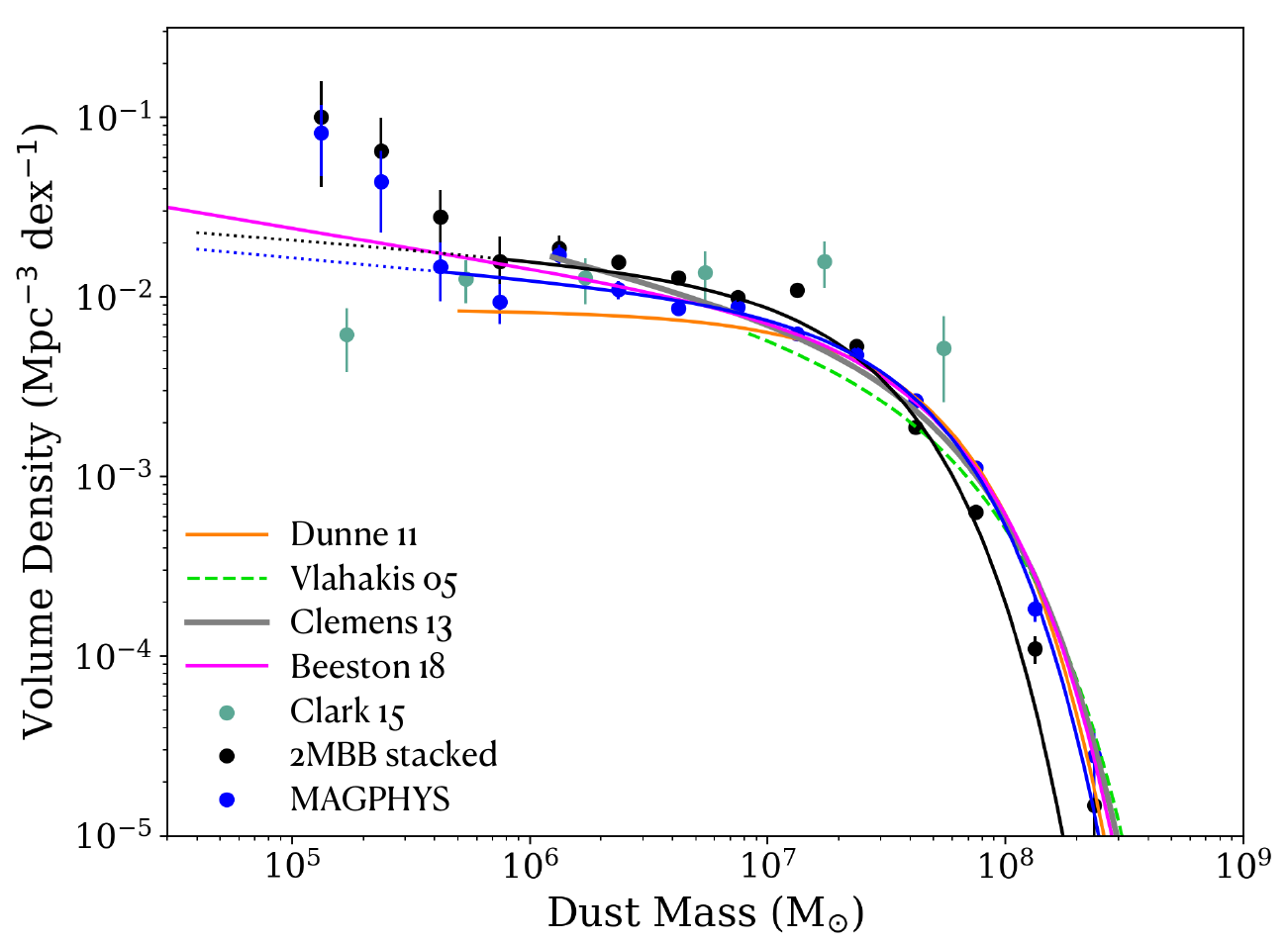}
 \includegraphics[trim=20mm 10mm 15mm 0mm clip=true,width=\columnwidth]{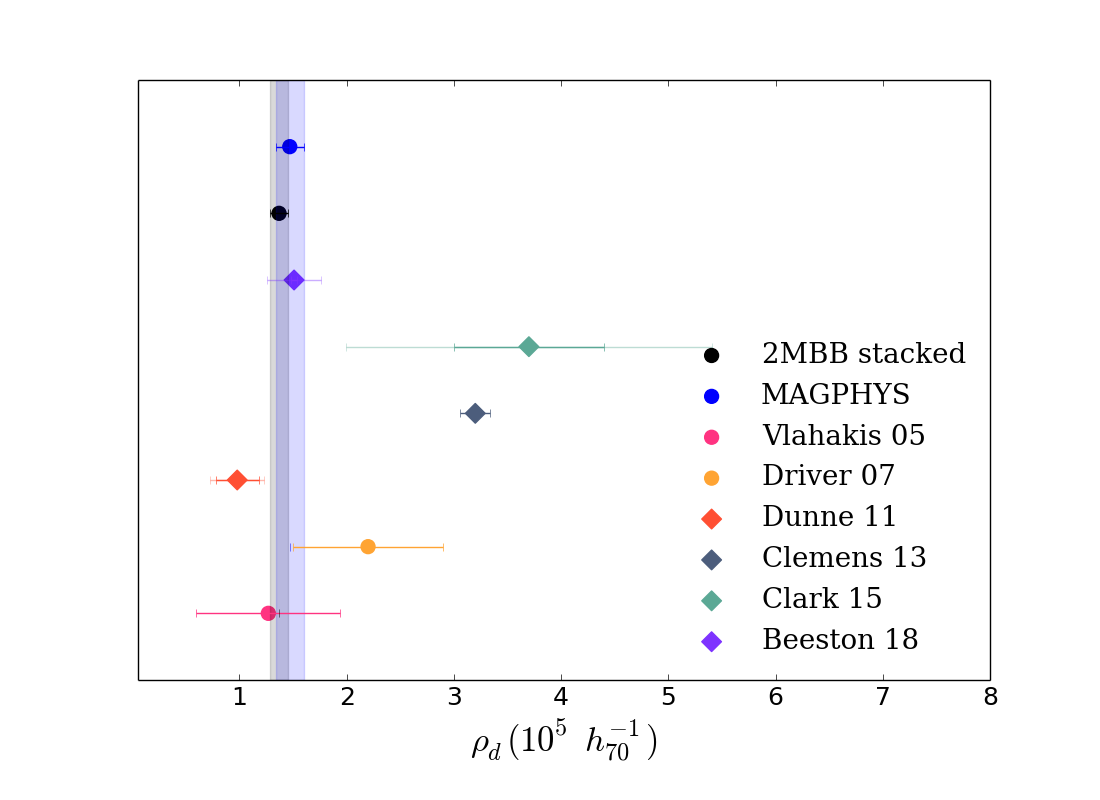}
 \caption{{\it Top:} Comparison of the low redshift ($z<0.1$) DMFs from this work (\magphys-based in blue, 2MBB stacked in black) with those from the literature.  We compare with (i) the blind, local $z<0.01$ galaxy sample from \citet{Clark2015} (ii) the all-sky local star-forming galaxies from the bright \textit{Planck} catalogue from \citet{Clemens2013} (iii) the ground-based submm measurements of local optical galaxies from \citet{Vlahakis2005} (iv) the 222 FIR-selected galaxies out to $z<0.1$ from the {\it H-}ATLAS survey \citep{Dunne2011} and (v) the $pV_{\rm max}$ DMF from the optically-selected galaxies in $H$-ATLAS from \citetalias{Beeston2018}.  Schechter fits to the data are shown by the solid lines. {\it Bottom:} Comparison of the low redshift dust mass density $\rho_d$ from this work with those from the literature. The dust density parameter measurements are scaled to the same cosmology, with diamonds representing dust-selected measurements, and circles representing optically-selected samples. The solid error bars indicate the published uncertainty whilst the transparent error bars indicate the total uncertainty derived by combining the published uncertainty and the cosmic variance uncertainty for that sample (where known). The shaded regions in black and purple emphasise the range of $\rho_d$ derived from this work with width showing the error from the combination of cosmic variance and statistical uncertainty.}
 \label{fig:dmf_others_C6}
\end{figure}

\begin{table*}
\centering
    \begin{tabular}{ccccccc}

     \multicolumn{5}{c}{Luminosity Function}\\
     \hline
  $z$   &   log$L^{*}$  	&		 $\alpha$  			& $\phi^{*}$  & \multicolumn{3}{c}{$\rho_{\rm L}$}   \\
 & ($h^2_{70}\,\rm W\,Hz^{-1}$) &    & ($10^{-3}\,h^3_{70}\,\rm Mpc^{-3}\,\rm dex^{-1}$) & \multicolumn{3}{c}{($10^{21}\rm W\,Hz^{-1}\, Mpc^{-3}$)}\\  
  & & & & Value & Error &  CV \\ \hline 
  $0.0 - 0.1$   &   24.11 $\pm$ 0.02  & -1.19 $\pm$0.03 & 1.42 $\pm$ 0.10 & 3.88 & $\pm$0.04 & $\pm$0.47\\ \\
    $0.1 - 0.2$ &   24.10 $\pm$ 0.01 &-1.19 & 4.10 $\pm$ 0.2 & 4.45& $\pm$0.05 & $\pm$0.55 \\ \\
   $0.2 - 0.3$   &  24.45 $\pm$ 0.01  & -1.19 & 1.27 $\pm$ 0.05 & 7.44 & $\pm$0.11 & $\pm$0.67 \\ \\
           $\mathit{0.3 - 0.4}$  &   $\mathit{24.61 \pm 0.01}$ & $\mathit{-1.19}$ & $\mathit{1.13 \pm 0.054}$ & $\mathit{9.6}$ & $\mathit{\pm 0.18}$ & $\mathit{\pm 2.21}$  \\ \\
           $\mathit{0.4 - 0.5}$   &  $\mathit{24.69\pm0.01}$  & $\mathit{-1.19}$ & $\mathit{1.30\pm0.07}$ & $\mathit{13.22}$& $\mathit{\pm 0.4}$  & $\mathit{\pm 4.23}$  \\ \hline
    \end{tabular}
        \caption{Best-fitting Schechter function values for luminosity functions derived in five redshift bins for our sample with the PC00 estimator. Uncertainty estimates are derived from a bootstrap analysis whereby 1000 realisations of the LF are fitted and the variance determines the uncertainty on each SF parameter.  The Error column indicates the error derived from the bootstrap analysis, and the CV column highlights the uncertainty due to cosmic variance. We have listed the highest redshift bins in italics to acknowledge the poorer sampling of galaxies in these bins at the knee of the function.   } 
    \label{tab:schechterTabLF}
\end{table*}

\begin{table*}
\centering
    \begin{tabular}{ccccc} 
    \multicolumn{5}{c}{Dust Mass Function $z < 0.1$}\\
    \hline 
    Survey   			&		 	log	$M_d^{*}$  	& 		 $\alpha$  			& $\phi^{*}$  &  $\rho_d$  \\
		& ($10^7\,h^2_{70}\,\rm M_{\odot}$) &    & ($10^{-3}\,h^3_{70}\,\rm Mpc^{-3}\,\rm dex^{-1}$) 
		& ($10^{5}\,h^{-1}_{70} \,\rm M_{\odot}\,Mpc^{-3}$)\\ \hline
		{\em Optically-selected} & & & & \\
    \citetalias{Beeston2018} $V_{\rm max}$ & 4.65$\pm$0.18 &  -1.22$\pm$0.01 & 6.26$\pm$0.28
    & 1.51$\pm$0.03 \\
    \citetalias{Beeston2018} BBD & 4.67$\pm$0.15 &  -1.27$\pm$0.01 & 5.65$\pm$0.23 &
    1.51$\pm$0.03 \\ \hline
    {\em FIR-selected} & & & & \\
    This work $V_{\rm max}$ \magphys  &   7.58$\pm$0.02  & -1.12$\pm$0.04 & 1.68$\pm$0.12 &
    1.26$\pm$0.09 \\
   PC00 \magphys    &   3.82$\pm$0.20  & -1.15$\pm$0.03 & 8.18$\pm$0.56 & 
   1.47$\pm$0.13 \\ \\

     This work $V_{\rm max}$ 2MBB stacked  & 7.43$\pm 0.01$  & -1.11$\pm$0.01 & 2.42$\pm$0.07 & 1.03$\pm$0.03\\
    PC00 2MBB stacked & 2.55$\pm$0.09  & -1.11$\pm$0.04 & 11.58$\pm$0.53 & 1.37$\pm$0.08\\ \hline
    \end{tabular}
        \caption{Best-fitting Schechter function values for the dust mass functions from the optically selected sample from \citetalias{Beeston2018} and the FIR-selected sample in this work over the same area of the sky at $z<0.1$. \citetalias{Beeston2018} used two different estimators $V_{\rm max}$ and BBD (the Bivariate Brightness Distribution, \citealp{Wright2017}). Here we compare both the estimators for the statistical functions ($V_{\rm max}$ and PC00) and the two dust mass estimates (\magphys~and two-temperature component MBB fits to the stacked luminosities).  Errors in $\rho_d$ are derived from the fits to the DMF and do not include cosmic variance. }
    \label{tab:schechterTab_C6}
\end{table*}

We fit a function to the luminosity density as a function of redshift, $\rho_{\rm L} \propto (1+z)^n$. Errors are estimated by randomly perturbing the data within the individual error (the combined statistical error from the fit parameters and the cosmic variance) and refitting the new samples. We find $n=6.80\pm 2.38$ provides a good fit using the PC00 estimator (with a similar result from $V_{\rm max}$). This is in agreement with the relationship found by \cite{Dye2010} for the 250-$\mu$m selected sources from the 14\,deg$^2$ $H$-ATLAS SDP field out to $z=0.2$ ($n=7.1$) and by \citet{Saunders1990} for 60\,$\mu$m sources out to $z=0.25$ ($n=6.7$).

The evolution seen in the LF can either be driven by the properties of the dust present at different cosmic times, or by the amount of dust in galaxies at different epochs. If the evolution of the LF was due to an increase in dust temperature due to increased star formation activity (star formation is known to peak around a redshift of 2), then we would expect either the stacking analysis or the \magphys~results to display a tendency for temperature to increase rapidly with redshift. Although an increase in $T_d$ was seen in the SEDs stacked in $L-z$ bins, the increase is only on the order of $\sim$1\,K in mass-weighted temperature in the redshift range $0.1 < z < 0.5$. Such a small temperature change would result in a change in luminosity\footnote{adopting a classical $L_{\rm IR} \propto M_{\rm dust}T^{(4+\beta)}$
and dust emissivity index $\beta = 2.0$.} of $<1.5$, lower than the factor $\sim$ 2.5 observed in Fig.~\ref{fig:LF_PC00vVmax}.

\subsection{The Dust Mass Function}

\begin{figure}
 \includegraphics[trim=0mm 0mm 0mm 0mm clip=true,width=1\columnwidth]{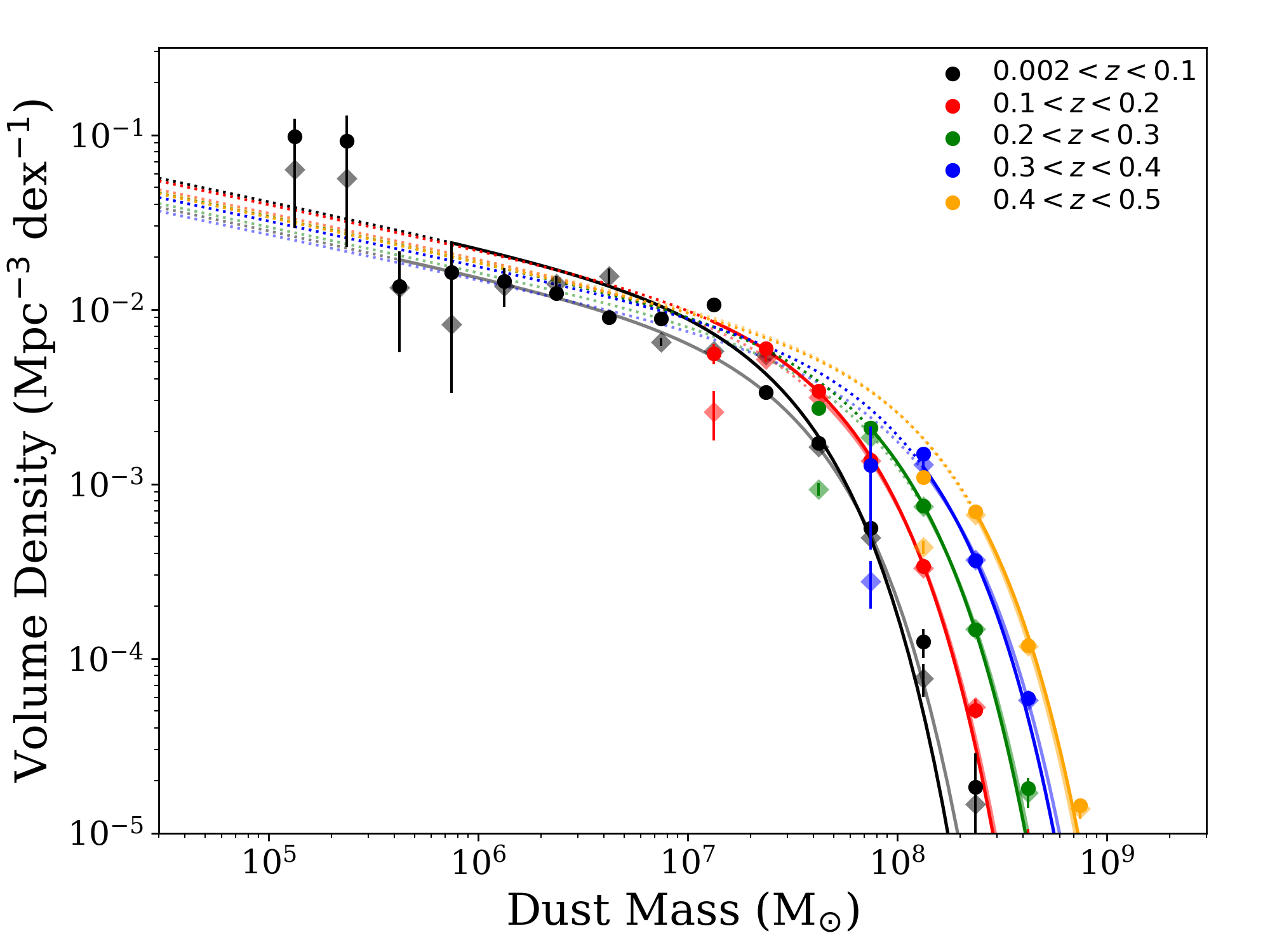}
   \includegraphics[trim=0mm 0mm 0mm 0mm clip=true, width=1\columnwidth]{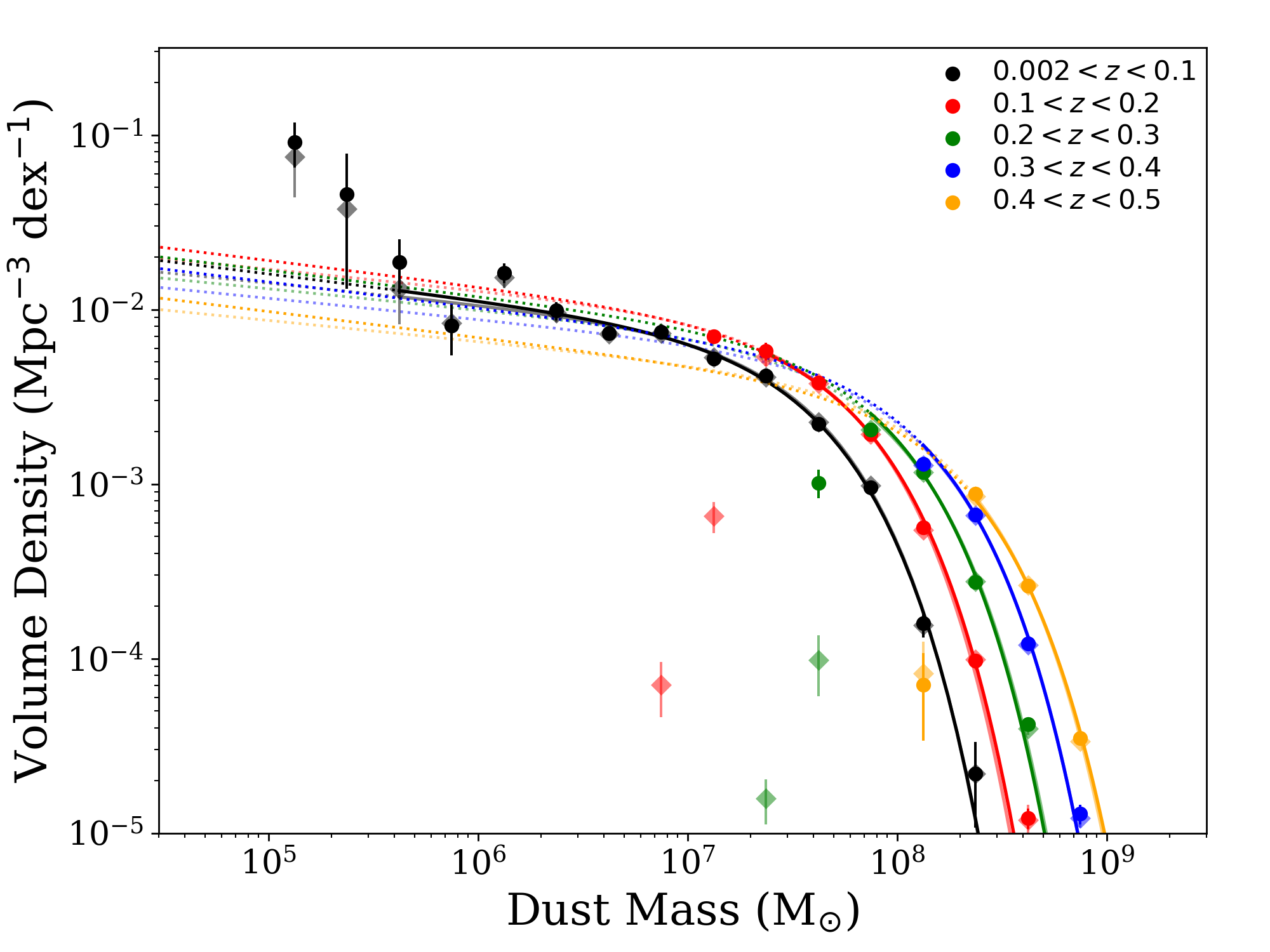}
 \caption{The DMF from ({\it top:}) the masses derived by the 2MBB stacking method, and ({\it bottom:}) the \magphys-based~dust masses produced using the $V_{\rm max}$ (transparent diamonds), and PC00 (opaque circles) estimators for 5 redshift slices. SF fits to the PC00 and $V_{\rm max}$ DMFs for each redshift slice are shown as solid and dashed curves respectively. Error bars are derived from a bootstrap analysis. }
 \label{fig:dmf_PC00vVmax}
\end{figure}

\subsubsection{The Low redshift dust mass function}
\label{DMFshape}

Many studies of the DMF focus on the local Universe since until the launch of \herschel, it was difficult to observe large areas of sky to a sufficient depth to measure redshift evolution. Here we compare the lowest redshift slice DMF from this work to the literature.  Although the low $z$ DMF covers the same redshift range as \citetalias{Beeston2018} and \citetalias{Dunne2011}, this work represents the largest FIR-selected sample used to derive a DMF. It is well described by a Schechter fit (Figure \ref{fig:dmf_others_C6} with fit parameters provided in Table \ref{tab:schechterTab_C6}).  The integrated dust mass density ($\rho_{\rm d}$) is calculated using the incomplete gamma function to integrate the DMF down to $M_{d} = 10^{4}\,M_{\odot}$, in line with \citetalias{Beeston2018}. We find $\rho_{\rm d} = (1.37 \,[1.47] \pm 0.08) \times 10^5\,\rm M_{\odot}\,Mpc^{-3}$ and dust mass density parameter
$\Omega_d = (1.01 \,[1.08] \pm 0.06) \times 10^{-6}$ for the stacked 2MBB results [\magphys~results]\footnote{To determine the dust density parameter $\Omega_d$, divide $\rho_{\rm d}$ by the critical density at $z=0$ where $\rho_{c,0} = 1.36\times 10^{11}\,\rm M_{\odot}\,Mpc^{-3}$ for our assumed cosmology.}. The integrated dust density parameter
corresponds to an overall fraction of baryons
(by mass) stored in dust $f_{\rm mb}
({\rm dust}) = (2.22\pm 0.13) \times 10^{-5}$, assuming the Planck baryonic density parameter of $45.51 \times 10^{-3}h^{-2}_{70}$ \citep{Planck2016cosmo}. The dust density determined here is 10\,per\,cent lower than the optically selected DMF from \citetalias{Beeston2018} and 30\,per\,cent higher than \citetalias{Dunne2011}. There is no significant offset between the different estimators/surveys in the total integrated dust mass at low redshifts, aside from the \citet{Clemens2010} and \citet{Clark2015} studies, where the latter has the largest uncertainty due to its small volume (larger cosmic variance, Figure \ref{fig:dmf_others_C6}, bottom). 

Our stacking analysis produces a DMF which has a different shape to previous work with more low-dust mass galaxies and fewer high-dust mass galaxies (and higher $\phi*$ with lower $M*$). Despite this, the dust density parameter is broadly consistent with literature values. The DMF obtained from the \magphys-based dust masses is in closer agreement to the previous studies particularly in the low and high mass ranges. We see no statistically significant offset between the dust content of galaxies in a large optically-selected sample (\citetalias{Beeston2018}) and the FIR-selected sample in this work, contrary to \cite{Clark2015} who found that in the nearby Universe ($z<0.01$) FIR-selected surveys are much more sensitive to colder, dust-rich galaxies. We also note that their work suffered from high cosmic variance errors and small numbers, but they also suffered fewer selection effects due to the low volume probed ($z<0.06$); they were more sensitive to cold dust galaxies than we are in this work. We also note that the highest dust mass bin of the 2MBB stacked mass DMF appears to be slightly underestimated by our SF fit. It is possible that the dust properties derived through stacking for each $L-z$ bin may be a useful probe of general trends, but when applied to individual galaxies or bins with small numbers of galaxies, this method may not be appropriate for estimating physical properties. 

\subsubsection{Evolution of the Dust Mass Function}

\begin{figure}
\centering
 \includegraphics[trim=0mm 8mm 2mm 0mm clip=true,width=1.05\columnwidth]{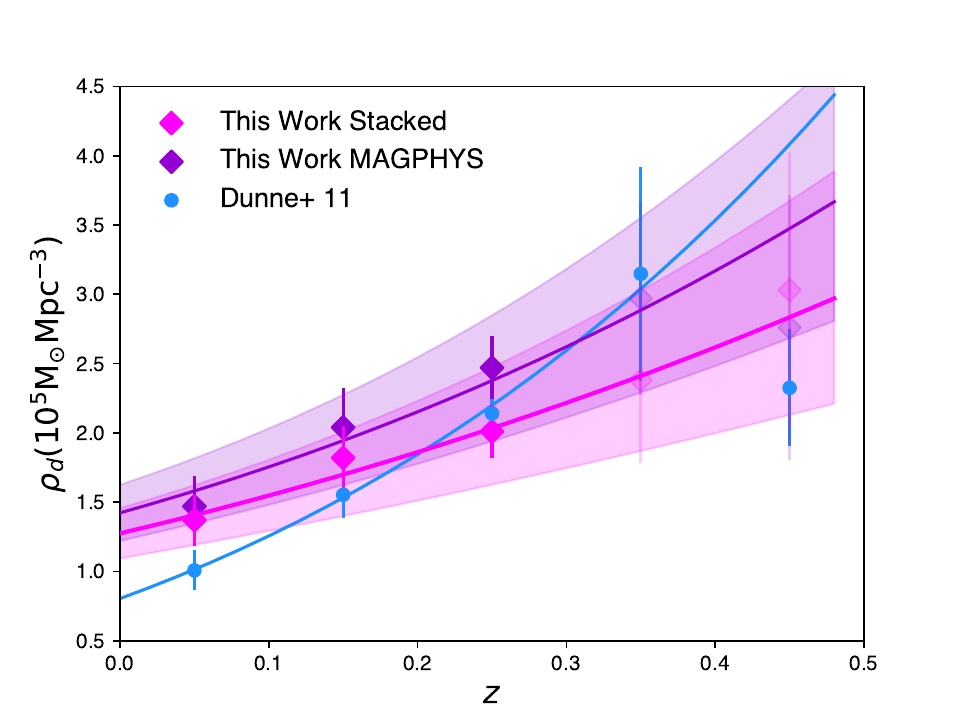}
 \caption{The dust mass density $\rho_{\rm d}$ as a function of redshift derived in this work for the stacked dust masses {\it (magenta diamonds)} and the \magphys\ dust masses {\it (purple diamonds)} with the highest redshift bins shown as transparent markers to acknowledge the poorer sampling of galaxies in these bins at the knee of the function and at lower masses. Error bars are derived from combining the errors from the DMF fit and the cosmic variance at each redshift. The \citetalias{Dunne2011} results corrected for our cosmology are shown in blue. The solid lines indicate the best fits $\rho_{\rm d} \propto (1+z)^n$ to each dataset. The shaded regions indicate the 1-$\sigma$ spread in the best-fit power law values. }
 \label{fig:evolution}
\end{figure}

\begin{figure*}
    \centering
    \includegraphics[trim=2mm 10mm 2mm 0mm clip=true,width=\textwidth]{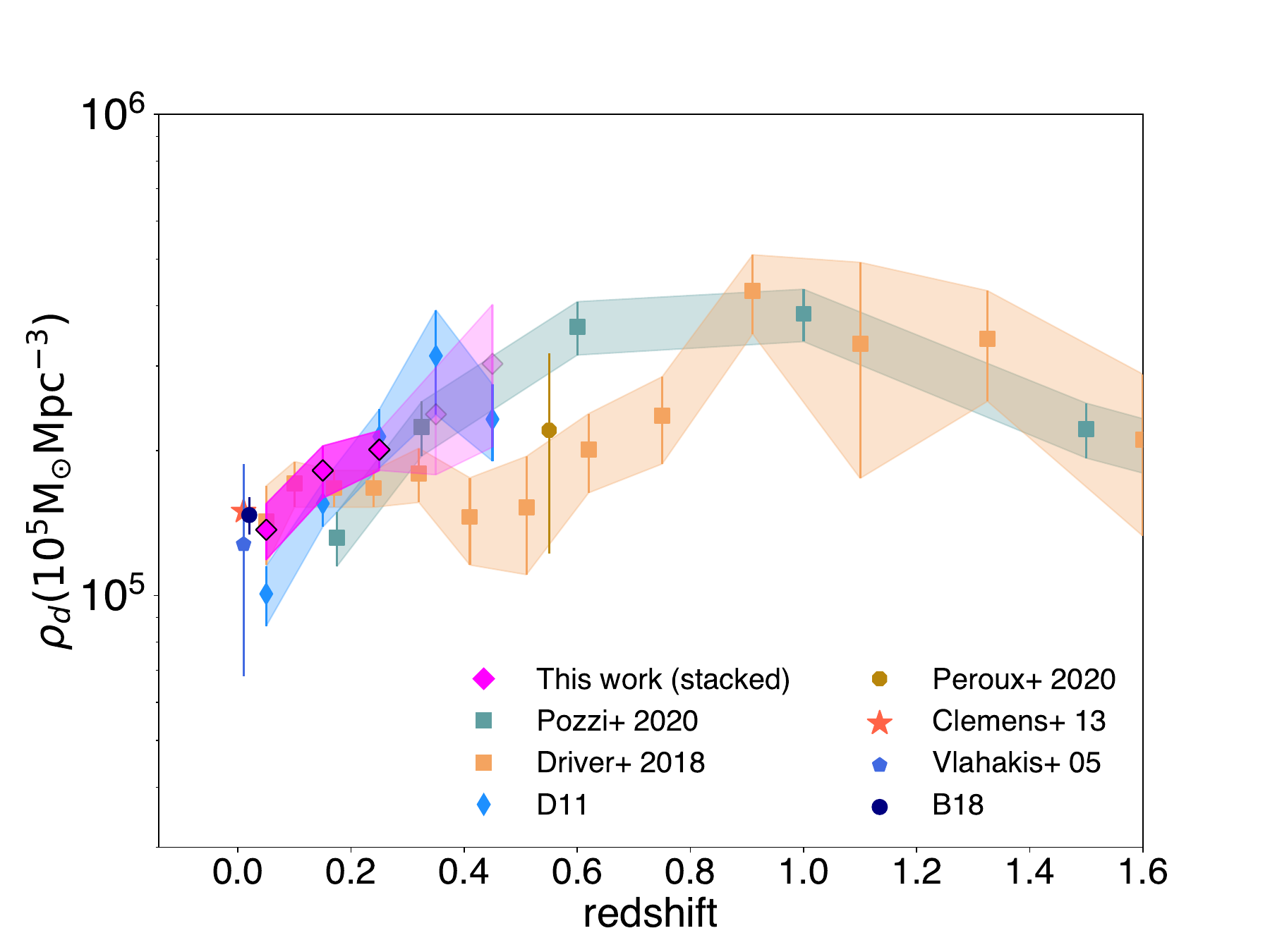}
    \caption{We compare the evolution of dust mass density to several literature estimates which have all been scaled to our assumed cosmology including \citet{Vlahakis2005}, \citetalias{Dunne2011}, \citet{Clemens2013}, \citetalias{Beeston2018}, \citet{Driver2018}, \citet{Pozzi2020}, \citet{Peroux2020}, with statistical errors shown (see legend for details). The shaded region shown for this work (magenta) is the uncertainty derived by combining the errors from the dust mass function and the cosmic variance. The \citet{Driver2018}, \citet{Peroux2020} and \citet{Pozzi2020} data values are taken from from the review paper of \citet{Peroux2020} (their Figure 12, supplementary Table 6). The dust mass densities shown here assume the same dust absorption coefficient factor, $\kappa_{\lambda}$ for \citetalias{Dunne2011}, \citetalias{Beeston2018}, \citet{Driver2018}, \citet{Clemens2013} and \citet{Vlahakis2005}.  \citet{Pozzi2020} instead assumes a $\kappa_{\lambda}$ that is a factor of 2 lower than the others, yielding dust mass densities a factor of 2 higher than their published numbers if scaled to this work. }
    \label{fig:dust_evolution_lit}
\end{figure*}

The DMFs for our FIR selected sample of 29,241 galaxies are split into 5 redshift bins (Figure \ref{fig:dmf_PC00vVmax}), with the best-fitting SF parameters for each redshift bin listed in Table \ref{tab:schechterTabDMF}. As with the LF, here we set $\alpha$ in the high redshift bins to the value derived for the $z<0.1$ bin. Here we can see the difference in the results of the PC00 method compared to $V_{\rm max}$ more obviously than in the luminosity function: the downturn in the lowest mass bins in each redshift slice is clearly visible in the bottom panel of Figure \ref{fig:dmf_PC00vVmax}.  (The difference is even more obvious in the dust mass function generated using the \magphys\ fits.)  For the remainder of our analysis we will discuss only the PC00 estimates of the LF and DMF unless otherwise stated.

The SF to the dust mass function based on the larger sample in this work (Figure \ref{fig:dmf_PC00vVmax}) demonstrates that the DMF does strongly evolve out to redshift 0.3 (over the past 3.5 billion years). The characteristic dust masses ($M_d^*$) change by up to 0.7\,dex depending on the estimators used to derive the DMF and the method of deriving dust masses. Although the DMF appears to continue to evolve out to a redshift of 0.5, the poorer sampling of the DMF around the knee and at lower dust masses will lead to a greater extrapolation uncertainty in the highest redshift bins. 

While the increase in $M_d^*$ with redshift is apparent, it is also clear that there is a trend for $\phi^*$ to decrease with redshift. As $M^*$ and $\phi^*$ are correlated, the dust mass density $\rho_d$ (or the integrated dust density parameter, $\Omega_d$) in each redshift slice \textit{is a more robust measure of the evolution of the dust content of galaxies than the SF parameters}.  The evolution of the dust mass density with redshift is shown in Figure \ref{fig:evolution}, derived from integrating the dust mass functions down to $M_d = 10^{4}\,M_{\odot}$ (\citetalias{Beeston2018}). Note that the derived dust mass density $\rho_m$ is not sensitive to whether $\alpha$ is fixed at the low $z$ bin or left free to vary. This is due to the bulk of the dust mass density residing in the higher mass end of the DMF. The derived dust mass density is however sensitive to the dust mass limit of the integral in the highest redshift bin, due to the poorer sampling around the knee. Setting a lower limit of $M_d = 10^{8}\,M_{\odot}$ for the dust density integral for $0.4 < z < 0.5$ reduces the dust mass density by a factor of 1.37.

Figure \ref{fig:evolution} demonstrates clear evidence for evolution in the dust mass density in galaxies out to redshift 0.3 regardless of which method is used to estimate the dust mass. We fit the relation $\rho_d \propto (1+z)^n$ to describe the evolution of the dust density with redshift. We find $n=2.53\pm0.62$ and $n=3.00\pm0.58$ for the 2MBB stacked method and \magphys-based method of dust mass estimates respectively over the redshift range $0 < z \le 0.5$. Errors are estimated by randomly perturbing the data within the individual error in $\rho_d$ (the combined statistical error from the fit parameters and the cosmic variance) and refitting the new samples. The evolution appears to be stronger in the \magphys-based DMFs.  We attribute the difference in these two values to the lack of evolution in the mass-weighted dust temperatures in the \magphys-based method compared to the 2MBB stacked temperatures. In their smaller, FIR-selected, sample, \citetalias{Dunne2011} found that $n=4$ was a good representation of all but their final data point at $z=0.5$. Their value is higher than we find using either the 2MBB stacked masses, or the \magphys-based dust masses, the latter of which followed the same method \citetalias{Dunne2011} use to derive their dust densities (and for which we find the same linear relationship between $L_{\rm 250}-M_d$). At $0.1<z<0.3$ we see relatively good agreement between this work and \citetalias{Dunne2011} for the 2MBB stacked method, though the dust densities here are systematically higher than \citetalias{Dunne2011} in this redshift range. In their highest redshift bin, \citetalias{Dunne2011} saw a sharp drop in dust density. This was suggested at the time to be a result of the declining fraction of spectroscopic redshifts available at $z>0.35$. We see a similar dip in our highest redshift bin using the \magphys-based dust masses, but not for our 2MBB stacked masses which produces lower dust densities at redshifts $0.2 < z < 0.4$. The volume density in this bin is significantly lower in the \magphys~sample compared to the stacked results, this is true for both the $p{\rm V_{max}}$ and PC00 estimators. This could be due to the smaller sample of individual galaxies with \magphys~fits and/or due to the lower fraction of galaxies with a spectroscopic redshift (Figure~\ref{tab:redshifts}). However, Figure~\ref{fig:evolution} implies that the dust densities derived from the larger sample used in this work are less affected by this incompleteness compared to \citetalias{Dunne2011}.

In Figure \ref{fig:dust_evolution_lit} we compare our dust mass densities as a function of redshift with \cite{Driver2018}, \citetalias{Dunne2011}, \citetalias{Beeston2018}, \citet{Dunne2003} and \citet{Pozzi2020} scaled to our assumed cosmology and $\kappa_{500}$ using data compiled by \citet{Peroux2020}. Over the same redshift range as the \citetalias{Dunne2011} analysis, \cite{Driver2018} found no evolution in $\rho_d$. Conversely \citet{Pozzi2020} and \citetalias{Dunne2011} instead found a rapid increase in the dust mass density of galaxies at low redshifts ($z < 0.7$). This work indicates that there is evolution in the dust mass density in this redshift regime, though the characteristic dust mass $M_d^*$ we find here evolves by at most a factor of 3 between $0<z<0.5$ compared to the factor of 8 quoted in \citetalias{Dunne2011}. This result is robust even if we 
discount the two (poorer-sampled) highest redshift bins. There are a further two things to note: (i) our Schechter fits to the 2MBB stacked DMF slightly underestimate the high dust mass end (and hence we may be underestimating the amount of evolution) and (ii) as mentioned earlier, the fit parameters $M_d^*$ and $\phi_*$ are correlated. For (i) we see that the \magphys~SFs do not suffer as large an `underfit' at high $M_d$ and yet there also, we only see a factor of two evolution in $M_d^*$, lower than in \citetalias{Dunne2011}. For (ii) we can look instead at the change in $\rho_d$, where we find the dust mass density, $\rho_d$, 5 billion years ago is twice as high as present day compared to the factor of 3 quoted in \citetalias{Dunne2011}. 

\section{Discussion}

\subsection{Caveats when deriving dust masses from stacking}
\label{sec:caveats}

\begin{figure}
 \includegraphics[trim=5mm 5mm 0mm 0mm clip=true,width=1.05\columnwidth]{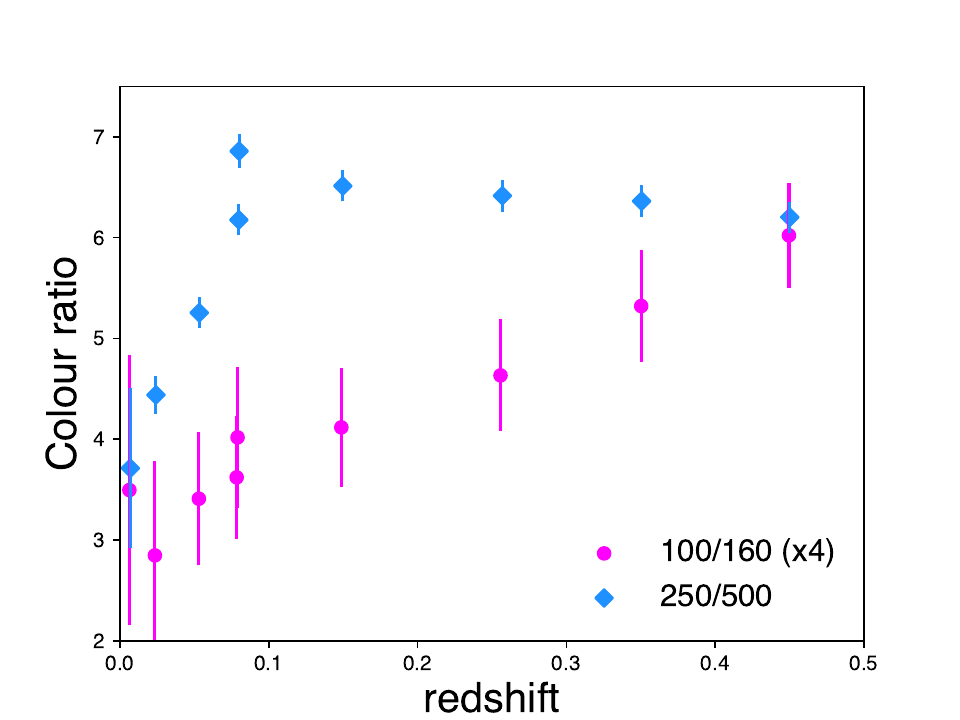}
 \caption{The ratio of 100 and 160\,\micron~luminosity (scaled by a factor of 4, {\it magenta circles}), and 250 and 500\,\micron~luminosity ({\it blue diamonds}) as a function of redshift for the stacked SEDs for our $L-z$ bins. The errors in each wavelength were added in quadrature. }
 \label{fig:colour_ratio_z}
\end{figure}

Whilst stacking may be a helpful tool in divining general trends between properties for a large number of galaxies, we note that the global properties of the stacked SED may not in fact be representative of all of the galaxies in the sample. Also, a single mass-weighted temperature could arise from very different SED shapes, and so we are assuming that the stacked SED is a good representation of the underlying dataset in terms of the cold and warm temperatures, as well as the fraction of the total luminosity assigned to each component. We simulated the effect of potential scatter in the mass-weighted temperatures produced by fitting a two-component MBB fit to a stacked SED by generating many pairs of temperatures and masses that produce given mass-weighted temperatures. We have found that even when stacking galaxy SEDs with the same mass-weighted temperature, the scatter in the resulting mass-weighted temperatures derived from the subsequent two component MBB fit can be around 3\,K.  As mentioned earlier, the sensitivity to cold dust is hampered at higher redshifts because of the intrinsic range of rest-frame frequencies probed. We could therefore be missing a whole class of galaxies from our analysis, either because they are simply too faint to be detected, or because only their warm dust is observable with our survey constraints.

\subsection{What other than mass could be driving the evolution in dust mass density with redshift?}

\subsubsection{Eddington Bias in the DMF?}

Here we check whether the scatter due to uncertainties in the dust masses of individual galaxies could introduce an Eddington bias into the DMF. This may occur if the underlying errors in dust mass scatters galaxies into neighbouring dust mass bins in either direction, combined with a non-uniform volume density across the mass bins. \citet{Loveday1992} showed that this
bias effectively convolves the underlying DMF with a Gaussian with width equal to the size of the scatter in the variable of interest (here dust mass) to give the observed DMF. Following \citetalias{Beeston2018}, we fit a Schechter function convolved with a Gaussian to the DMFs, where we estimate the width of the Gaussian using the mean error around the knee of the function in the different redshift bins. (We use the \magphys~derived DMF in order to test for this bias since the errors in dust mass are larger than the 2MBB method.) The mean error is estimated from half the difference between the 84th and 16th dust mass percentiles derived from \magphys, these correspond to 1$\sigma$ if the uncertainties are Gaussian. The errors are 0.13, 0.20, 0.20, 0.22, and 0.22\,dex. The resulting deconvolved Schechter functions have higher $\chi^2$ values in all redshift bins except for the first ($0 < z<0.1$). The deconvolved SFs still show evolution in the dust mass function, with redshift with the dust mass density, $\rho_d$ increasing by a factor of 2.4 [3.6] over $0 < z < 0.3$ [$0 < z < 0.5$]. In terms of the characteristic dust mass, the evolution is reduced by 30\,per\,cent in the redshift range $0 < z < 0.3$, but the evolution is similar over $0 < z < 0.5$. 

\subsubsection{Evolution in dust temperature?}
Using the relationship $M_{d} \propto T^{-2.4}$ (\citetalias{Dunne2011}), we can estimate the change in temperature required in order for the evolution of $\rho_d$ to be flat across our $z$ bins (i.e. equal to the low-redshift dust density). For the 2MBB stacked masses, the mass-weighted temperature at each of the higher redshift slices in turn would have to be 23.2\,K, 24.7\,K, 26.7\,K, and 29.8\,K. For the \magphys-based masses we found that all redshift slices have a mass-weighted temperature of 20\,K. This may be due to the possibility that \magphys~is returning the prior dust temperature in faint FIR sources; this bias would act to flatten any underlying trend in dust temperature with redshift, thereby enhancing the evolution in $\rho_d$.  In order for there to be no evolution in $\rho_d$ with redshift, the \magphys-based mass-weighted temperatures would need to be 23.0\,K, 24.8\,K, 26.9\,K, and 26.1\,K in each redshift slice.  Although we do observe some evolution in dust temperature with $z$ in our stacked SEDs, we require a much stronger evolution with temperature than we observe in both \magphys~and stacking in order for the change in dust density with redshift to be explained via dust temperature alone.

\subsubsection{Possible bias in the submillimetre colours of \herschel~sources?}

It is also possible that any evolution in temperature with redshift could be underestimated if the effect of confusion on the flux measurements in the $H$-ATLAS catalogue are not properly accounted for. Contamination can also arise because of galaxy lensing (e.g. \citealt{Negrello2017}), whereby the light from a higher redshift source can be deflected by an intermediate redshift one towards the observer. A typical dust SED with temperatures around 20\,K will peak at around 120\,\micron~in the observed frame at $z=0.1$, but the same source at $z=0.5$ would peak at 180\,\micron~in the observed frame. This means that contamination from high redshift sources gets stronger with longer wavelengths, compounded by the increasing \herschel~beam size. Recent work by \citet{Dunne2020_lens} using ALMA and  \herschel\ data has shown that this effect could represent up to 13, 26, and 44\,per\,cent of the flux contribution of the measured flux at 250, 350, and 500\,\micron~respectively\footnote{Lower values of this effect were originally estimated using the $H$-ATLAS maps alone \citep{Valiante2016}.} at $z \sim 0.35$. The fraction of contamination could evolve with redshift, since the probability that a galaxy will be a lens peaks around $z\sim 0.3-0.4$, and at lower redshifts galaxies are unlikely to be affected by lensing (e.g. \citealt{Ofek2003}). It is possible then that the higher redshift bins may be biased to lower dust temperatures due to the artificial increase in the flux at longer wavelengths from contaminating lensed sources. We can probe this by looking at the FIR colours with redshift. The 100-to-160\micron~rest-frame luminosity ratio is seen to increase with redshift (Figure \ref{fig:colour_ratio_z}), but the same is not true for the 250 to 500\,\micron~ratio beyond a redshift of 0.1.  The increase we observe in mass- and luminosity-weighted dust temperatures in the higher redshift slices therefore appear real, but the effect could be somewhat underestimated by our stacking process. A larger sample of galaxies with high quality spectroscopic redshifts and FIR measurements is needed to confirm the \citet{Dunne2020_lens} result.


\begin{table*}
\centering
    \begin{tabular}{cccccccc}
    \hline \hline
 Method &  $z$   &   log$M^{*}$  	&		 $\alpha$  			& $\phi^{*}$  & \multicolumn{3}{c}{$\rho_d$}   \\
 	&	& ($h^2_{70}\,\rm M_{\odot}$) &    & ($10^{-2}\,h^3_{70}\,\rm Mpc^{-3}\,\rm dex^{-1}$) & \multicolumn{3}{c}{ ($10^{5}\,h^{-1}_{70} \rm \,M_{\odot}\,Mpc^{-3}$) }  \\ 
 	&	&  &    & & Value & Error & CV  \\ \hline \hline
         &  $0.0 - 0.1$   &    2.55$\pm$0.09  & -1.11$\pm$0.04 & 11.58$\pm$0.53 & 1.37 & $\pm$0.08 &  $\pm$0.18  \\ \\
         &  $0.1 - 0.2$   &   7.61$\pm^{0.01}_{0.01}$  & -1.11 & 2.28$\pm$0.07 & 1.82 & $\pm$0.06 & $\pm$0.25    \\ \\
    PC00 2MBB stacked    &  $0.2 - 0.3$   &  7.8$\pm^{0.01}_{0.01}$  & -1.11 & 1.63$\pm$0.05 & 2.01 & $\pm$0.06  &  $\pm$0.22 \\ \\
         &  $\mathit{0.3 - 0.4}$   &  $\mathit{7.95\pm^{0.02}_{0.02}}$  & $\mathit{-1.11}$ & $\mathit{1.37 \pm 0.14}$ & $\mathit{2.38}$ & $\mathit{\pm 0.24}$ &  $\pm$$\mathit{0.68}$  \\ \\
         &  $\mathit{0.4 - 0.5}$   &  $\mathit{8.07\pm^{0.01}_{0.01}}$  & $\mathit{-1.11}$ & $\mathit{1.32 \pm0.10}$ & $\mathit{3.03}$ & $\mathit{\pm 0.23}$ &$\pm$$\mathit{0.88}$ \\ \\ \hline
         &  $0.0 - 0.1$   &  3.82$\pm$0.20  & -1.15$\pm$0.03 & 8.18$\pm$0.56 & 1.47 & $\pm$0.13   \\ \\
         &  $0.1 - 0.2$   &   7.58$\pm^{0.02}_{0.02}$  & -1.15 & 4.70$\pm$0.32 & 2.04 & $\pm$0.14   \\  \\
    PC00  \magphys   &  $0.2 - 0.3$   &   7.91$\pm^{0.01}_{0.01}$  & -1.15 & 2.46$\pm$0.04 &  2.47 & $\pm$0.04   \\ \\
        & $\mathit{0.3 - 0.4}$  &   $\mathit{8.09\pm^{0.01}_{0.01}}$  & $\mathit{-1.15}$ & $\mathit{1.20 \pm 0.05}$ & $\mathit{2.97}$ & $\mathit{\pm 0.12}$   \\ \\
         &  $\mathit{0.4 - 0.5}$   &   $\mathit{8.25\pm^{0.02}_{0.02}}$  & $\mathit{-1.15}$ & $\mathit{0.77\pm0.10}$ & $\mathit{2.76}$ & $\mathit{\pm 0.36}$  \\  \hline
    \end{tabular}
        \caption{Best-fitting Schechter function values for the DMFs derived in five redshift bins using the PC00 estimator and two methods for determining dust masses (\magphys~and stacking). The Error column indicates the error derived from a bootstrap analysis, and the CV column highlights the uncertainty due to cosmic variance. The corresponding values for the $V_{\rm max}$ estimator are shown in Table~\ref{tab:schechterTabDMF_Vmax}. We include the fraction of the sources with photometrically-derived redshifts. We have listed the highest redshift bins in italics to acknowledge the poorer sampling of galaxies at the knee of the function and the extrapolation to low masses based on the redshift zero bin - note that the fit errors quoted for these bins likely underestimate the uncertainty. }
    \label{tab:schechterTabDMF}
\end{table*}

\section{Conclusion}

We have been able to derive the dust density in redshift bins out to $z=0.5$ using a FIR-selected sample over a larger area of sky (factor 12) and with an order of magnitude more galaxies than used in previous analyses. We measure dust properties for 29,241 FIR-selected galaxies using two different methods. We find that: 

\begin{itemize}

\item the mass-weighted dust temperature appears to increase with luminosity at low redshifts when we use the two-temperature modified blackbody fits (2MBB) to the galaxies stacked in luminosity and redshift bins.

\item There is a strong increase in the luminosity-weighted temperature in the higher redshift slices. We are unable to determine whether this evolution was driven by increasing redshift, or simply due to the observed increase in luminosity with redshift.

\item There is a tendency for the \magphys~SED fitting routine to assign lower dust masses to low luminosity galaxies and higher dust masses to high luminosity galaxies compared to the two-temperature component MBB fits to the stacked luminosities.
\end{itemize}

We measure the Luminosity Function and Dust Mass Function in five redshift slices across our sample out to $z=0.5$. 

\begin{itemize}

\item We find reasonable agreement in the evolution with redshift of the luminosity with \cite{Dye2010} who performed a similar analysis using 1688 sources in the $H$-ATLAS SDP field. The evolution of the luminosity density in our sample can be fit by the relationship $\rho_{L} \propto (1+z)^{6.8 \pm 2.4}$ out to redshift 0.5.
 
\item We find that using either the \magphys-based masses or the 2MBB stacked masses, the dust density parameter evolves out to $z=0.3$ and tentatively out to $z=0.5$, and the evolution is stronger using the \magphys-based masses. The evolution of the dust density over the full redshift range is described by $\rho_d \propto (1+z)^{2.5\pm0.6}$ and $\rho_d \propto (1+z)^{3.0\pm0.6}$ for the 2MBB stacked masses and \magphys-based masses respectively. 

\item We attempt to account for Eddington Bias by fitting a deconvolved Schechter function to the dust mass volume density. In this scenario, the evolution of the dust mass density with redshift remains similar, but the evolution in the characteristic dust mass is reduced (to a factor of 2.4, from a factor of 3) in the range $0<z<0.3$. 

\item We show that the LF and DMF both evolve with redshift over the redshift range $0<z<0.3$ as originally seen in \citet{Dunne2011} and later in \citet{Pozzi2020}. We find clear evidence of evolution in the dust density over the past 3 Gyr, and tentative evidence that the evolution continues out to 5 Gyr. The evolution we derive is weaker than that found in the smaller FIR-selected survey of \citet{Dunne2011}. Due to the poorer sampling of the DMF in the highest redshift bins and the extrapolation required to derive the dust mass density, particularly in the redshift range $0.4 < z < 0.5$, we cannot rule out the possibility that the dust density evolution remains flat at these redshifts (\citealt{Driver2018}). Since the dust temperature does not evolve strongly enough with redshift to explain the observed evolution in the dust density of the Universe, we conclude that an increase in the dust mass content of the Universe over cosmic time is the driving force behind the evolution of the LF and DMF with redshift. 
\end{itemize}

\section*{Acknowledgements}
We thank the referee for their careful reading and very useful comments.
RAB, HLG, LD and SJM acknowledge support from the European Research Council (ERC) in the form of Consolidator Grant {\sc CosmicDust} (ERC-2014-CoG-647939). For the purpose of open access, the author has applied a Creative Commons Attribution (CC-BY) licence to any Author Accepted Manuscript version arising. SAE and MWLS acknowledge support from an STFC Consolidated grant. 
This research has made use of {\tt Astropy\footnote{http://www.astropy.org}}, a community-developed core Python package for Astronomy (\citealt{Astropy2013,Astropy2018}), and the Python libraries 
{\tt NumPy} (\citealt{Numpy}) and {\tt Matplotlib} (\citealt{Matplotlib}).

\section*{Data Availability}
The $H$-ATLAS and GAMA data used here can be obtained from \url{ https://www.h-atlas.org/public-data/download} and \url{ http://www.gama-survey.org/dr4/} respectively.



\bibliographystyle{mnras}
\bibliography{rosie} 




\appendix

\section{Comparison of redshift estimates}
\label{app:redshift}
\begin{figure}
\begin{center}
  \includegraphics[trim=0mm 10mm 0mm 0mm clip=true,width=1\columnwidth]{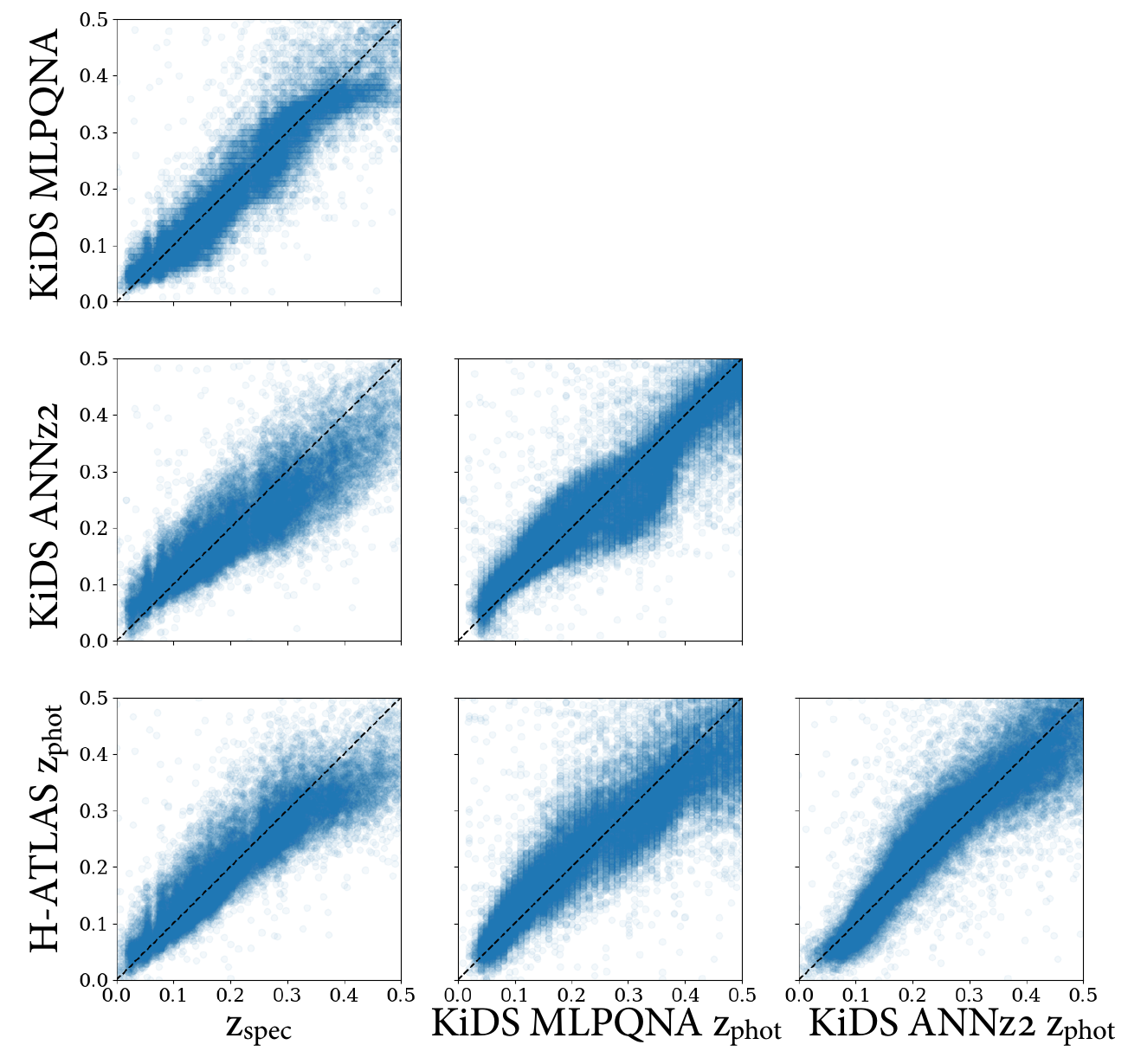}
 \caption{Grid comparing different redshift estimates for galaxies used in this work. Spectroscopic redshifts were collected from GAMA, SDSS DR7 and DR8, 2SLAQ LRG and QSO samples, 2dF, and 6dF. Three estimates of photometric redshifts are also shown, $H$-ATLAS using the ANNz software, and from KiDS, ANNz2 and MLPQNA. }
 \label{fig:z_compare}
 \end{center}
\end{figure}

The different redshift estimators available for our sample are compared in Figure~\ref{fig:z_compare} to check for any systematic biases that may be introduced in our estimates of dust mass.
MLPQNA is more tightly correlated with the GAMA spectroscopic redshifts, but since it is trained on relatively bright and nearby sources this is unsurprising.

\section{Estimating Dust Masses}
\label{app:fits}

\subsection{Blackbody Fits to the Stacked SEDs}
\label{app:blackbodyfits}
\begin{figure*}
 \includegraphics[trim=100 150 0 50 clip=true,width=0.42\textwidth]{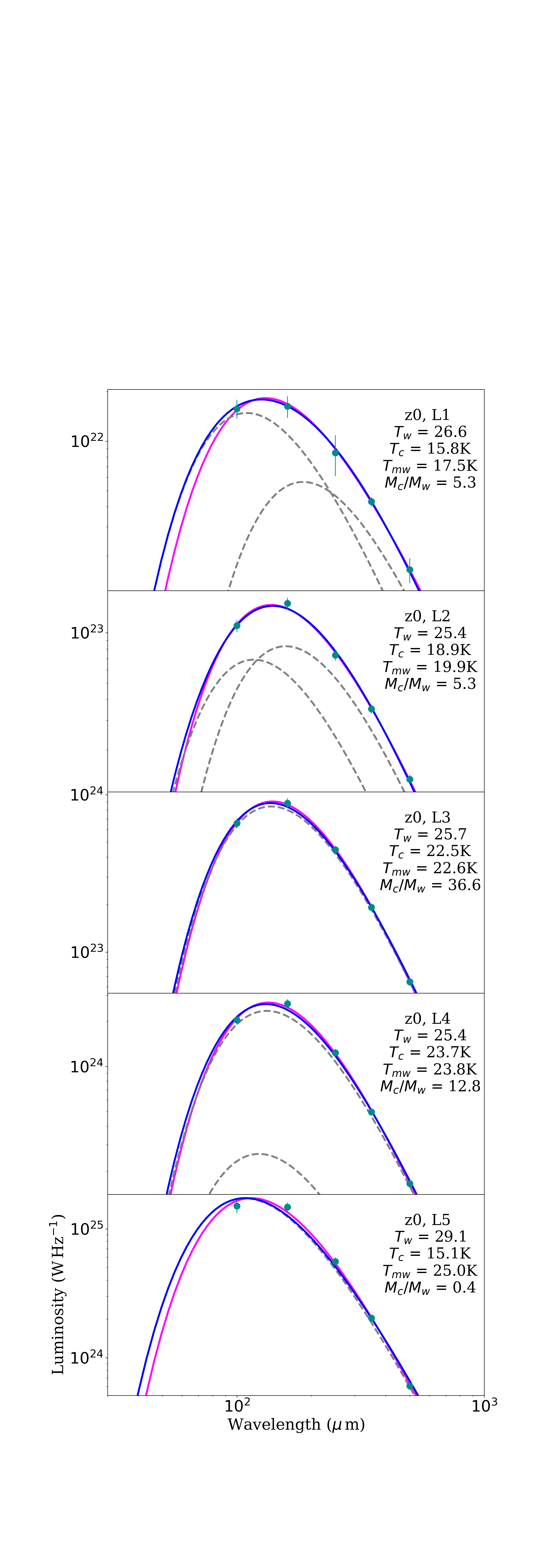}
  \includegraphics[trim=0 150 0 50 clip=true,width=0.47\textwidth]{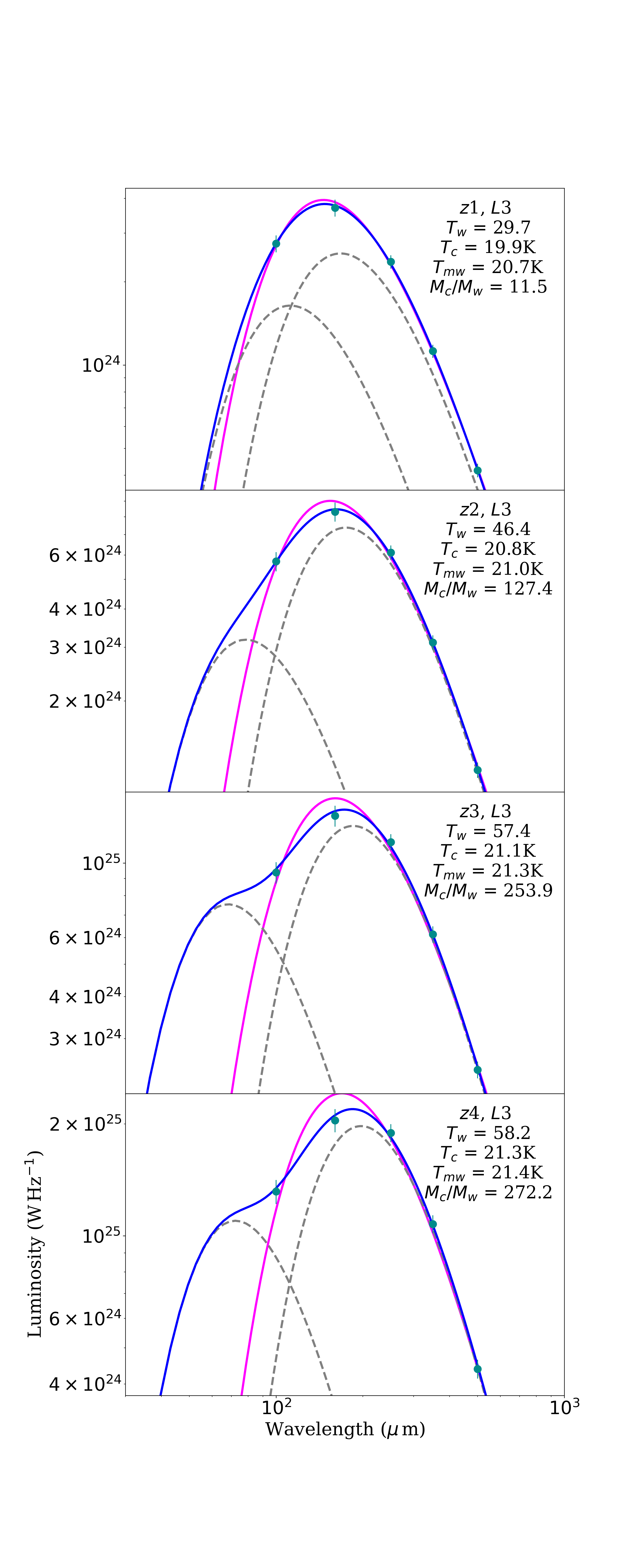}
 \caption{The stacked luminosity SEDs for the $L-z$ bins for {\it left:} the lowest redshift slice $z_0$ with increasing luminosity from top to bottom and {\it right:} for luminosity bin $L_3$ with increasing redshift from top to bottom. The bestfitting one and two temperature component MBB fits are shown in magenta and blue respectively. The dashed lines show the components of the two temperature fit,  $T_{\rm d, w}$ and  $T_{\rm d, c}$.}
 \label{fig:fitgrid1}
\end{figure*}

\begin{figure}
  \includegraphics[trim=0mm 10mm 0mm 0mm clip=true,width=\columnwidth]{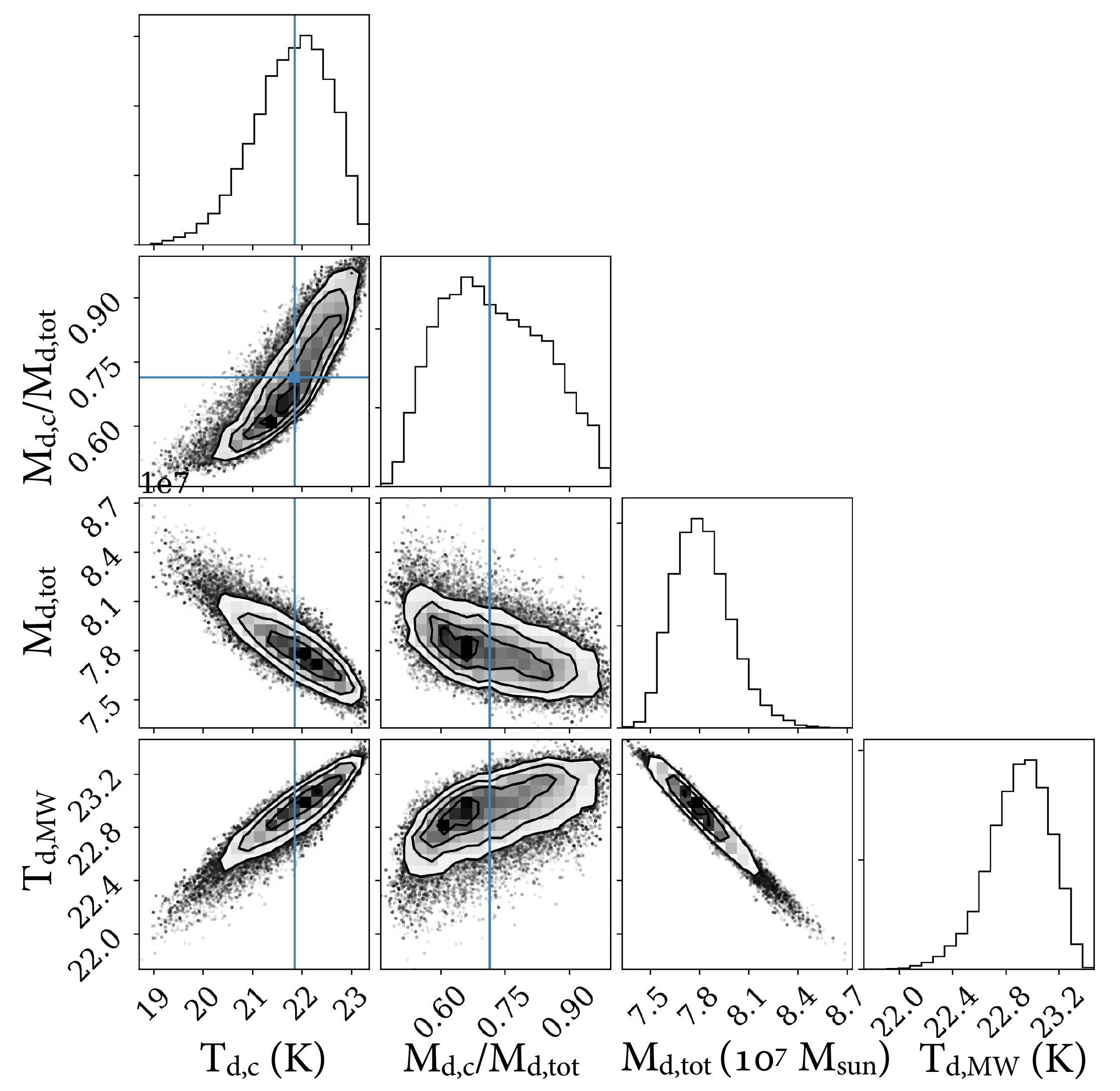}
 \caption{An example of the one and two dimensional projections of the probability distributions from fitting a two component MBB using MCMC sampling of the posterior for the highest $L-z$ luminosity bin in the lowest redshift slice ($L_5, z_0$). Shown are the total dust mass $M_{\rm d,tot}$ in $10^7 \,M_{\odot}$, mass-weighted dust temperature $T_{\rm d,MW}$ in K and the fraction of the dust mass in the cold component, i.e. $M_{\rm d,c}/M_{\rm tot}$.}
 \label{fig:corner}
\end{figure}

The fit properties from the one dust temperature component MBB fit to the stacked SEDs (1MBB) are provided in Table~\ref{tab:stacked_SED_params_one} with example SEDs shown in Figure~\ref{fig:fitgrid1}. An example of the MCMC results from the two component modified blackbody fits (2MBB) for the highest $L-z$ luminosity bin in the lowest redshift slice is provided in Figure~\ref{fig:corner}.

\begin{table*}
\centering
    \begin{tabular}{ccc|cccc} \hline
$\Delta z$ & $\Delta$log\,$L_{250}$  & $N$ & $T_d$ & log\,$M_d$ & $\beta$ & $\chi^2$  \\ 
           &  ($\rm W\,Hz^{-1}$)  &      &       (K)  & ($M_{\odot}$)            &  &  \\ \hline 
0.002 - 0.1 & 21.6 - 22.2 & 27 & 23.2$\pm$0.3 & 5.32$\pm$0.02 & 1.5$\pm$0.0 & 0.84  \\
0.002 - 0.1 & 22.2 - 22.9 & 167 & 22.4$\pm$0.2 & 6.38$\pm$0.01 & 1.5$\pm$0.0 & 1.64 \\
0.002 - 0.1 & 22.9 - 23.6 & 1147 & 24.1$\pm$0.4 & 7.05$\pm$0.01 & 1.5$\pm$0.0 & 0.18  \\
0.002 - 0.1 & 23.6 - 24.3 & 2201 & 23.3$\pm$0.4 & 7.43$\pm$0.01 & 1.8$\pm$0.0 & 0.17  \\
0.002 - 0.1 & 24.3 - 25.0 & 199 & 24.0$\pm$0.4 & 7.88$\pm$0.01 & 1.9$\pm$0.1 & 0.42  \\
0.1 - 0.2 & 23.7 - 25.3 & 7739 & 24.0$\pm$0.4 & 7.91$\pm$0.01 & 1.8$\pm$0.0 & 0.81 \\
0.2 - 0.3 & 24.2 - 25.4 & 6365 & 25.7$\pm$0.4 & 8.42$\pm$0.01 & 1.6$\pm$0.0 & 2.46 \\
0.3 - 0.4 & 24.5 - 26.1 & 5861 & 27.2$\pm$0.2 & 8.76$\pm$0.01 & 1.5$\pm$0.0 & 5.66  \\
0.4 - 0.5 & 24.7 - 25.8 & 5535 & 27.7$\pm$0.1 & 9.06$\pm$0.01 & 1.5$\pm$0.0 & 9.24  \\ \hline 
\end{tabular}
\caption{The best-fitting SED parameters derived for the stacked galaxies in each of our $L-z$ bins for the one temperature component MBB fit.}
\label{tab:stacked_SED_params_one}
\end{table*}

\subsection{The Dust Mass Function}
\label{app:dmf}
\begin{table*}
\centering
    \begin{tabular}{cccccc}
    \hline \hline
 method &  $z$   &   log$M^{*}$  	&		 $\alpha$  			& $\phi^{*}$  & $\rho_d$   \\ \\
 	&	& ($h^2_{70}\,\rm M_{\odot}$) &    & ($10^{-2}\,h^3_{70}\,\rm Mpc^{-3}\,\rm dex^{-1}$) & ($ 10^{5}\,h_{70}^{-1} \rm \,M_{\odot}\,Mpc^{-3}$)\\ \hline \hline
         &  $0.0 - 0.1$   &   7.43$\pm^{0.01}_{0.01}$  & -1.11$\pm$0.01 & 2.42$\pm$0.07 & 1.03$\pm$0.02 \\ \\
         &  $0.1 - 0.2$   & 7.64$\pm^{0.01}_{0.01}$ & -1.11& 1.97$\pm$0.05 & 1.66$\pm$0.02     \\ \\
    $V_{\rm max}$ 2MBB stacked    &  $0.2 - 0.3$ & 7.83$\pm^{0.11}_{0.16}$  & -1.11 & 1.34$\pm$0.41 & 1.78$\pm$0.54  \\ \\
         & $\mathit{0.3 - 0.4}$   &  $\mathit{8.00\pm^{0.02}_{0.02}}$  & $\mathit{-1.11}$ & $\mathit{1.10\pm 0.12}$ & $\mathit{2.13 \pm 0.13}$    \\ \\
         &  $\mathit{0.4 - 0.5}$   &  $\mathit{8.06\pm^{0.01}_{0.01}}$  & $\mathit{-1.11}$ & $\mathit{1.37\pm 0.05}$ & $\mathit{3.04 \pm 0.07}$     \\ \\ \hline
         &  $0.0 - 0.1$   &   7.58$\pm^{0.02}_{0.02}$  & -1.12$\pm$0.04 & 1.68$\pm$0.12 & 1.26$\pm$0.02     \\ \\
         &  $0.1 - 0.2$   &  7.73$\pm^{0.01}_{0.01}$  & -1.12 & 1.92$\pm$0.05 & 2.03$\pm$0.02    \\  \\
    $V_{\rm max}$ \magphys-based     &  $0.2 - 0.3$   &   7.92$\pm^{0.01}_{0.01}$  & -1.12 & 1.42$\pm$0.05 & 2.32$\pm$0.05  \\ \\
         &  $\mathit{0.3 - 0.4}$  &   $\mathit{8.09 \pm^{0.01}_{0.01}}$  & $\mathit{-1.12}$ & $\mathit{1.20\pm 0.07}$ & $\mathit{2.84 \pm 0.09 }$    \\ \\
         &  $\mathit{0.4 - 0.5}$   &   $\mathit{8.23\pm^{0.01}_{0.01}}$  & $\mathit{-1.12}$ & $\mathit{0.86\pm 0.05}$ & $\mathit{2.83\pm 0.13}$  \\ \\ \hline
 \hline
    \end{tabular}
        \caption{Best-fitting Schechter function values for the DMFs derived in five redshift bins using the $V_{\rm max}$ estimator and two methods for determining dust masses (\magphys~and stacking). Uncertainty estimates are derived from a bootstrap analysis. The uncertainty due to cosmic variance is listed in Table~\ref{tab:schechterTabDMF}. We have listed the highest redshift bins in italics to acknowledge the poorer sampling of galaxies in these bins at the knee of the function and the extrapolation to low masses based on the redshift zero bin - note that the errors quoted for these bins likely underestimate the uncertainty. } 
    \label{tab:schechterTabDMF_Vmax}
\end{table*}


\bsp	
\label{lastpage}
\end{document}